\documentclass[journal=jcisd8,manuscript=article]{achemso}
\usepackage{graphicx}
\usepackage{dcolumn}
\usepackage{bm}
\usepackage{amsmath}
\usepackage[version=4]{mhchem}

\usepackage{nameref}

\newcommand{\kbt}{$k_\mathrm{B} T$}
\newcommand{\kjmol}{ $\mathrm{kJ}\cdot \mathrm{mol}^{-1}$  }
\newcommand{\mkbt}{k_\mathrm{B} T}
\newcommand{\mkjmol}{ \mathrm{kJ}\cdot \mathrm{mol}^{-1} }

\newcommand{\progname}[1]{#1}

\title{NPCoronaPredict: A computational pipeline for the prediction of the nanoparticle - biomolecule corona}

\author{Ian Rouse}
  \email{ian.rouse@ucd.ie}
  \author{David Power}
    \author{Julia Subbotina}
\author{Vladimir Lobaskin}%

\affiliation{%
University College Dublin\\
 Belfield \\
 Dublin 4
}%
 \date{\today}
\begin{document}
\begin{abstract}
The corona of a nanoparticle immersed in a biological fluid is of key importance to its eventual fate and bioactivity in the environment or inside live tissues. It is critical to have insight into both the underlying bionano interactions and the corona composition to ensure biocompatibility of novel engineered nanomaterials. A prediction of these properties in silico requires the successful spanning of multiple orders of magnitude of both time and physical dimensions to produce results in a reasonable amount of time, necessitating the development of a multiscale modelling approach. Here, we present the \progname{NPCoronaPredict} open-source software package: a suite of software tools to enable this prediction for complex multi-component nanomaterials in essentially arbitrary biological fluids, or more generally any medium containing organic molecules. The package integrates several recent physics-based computational models and a library of both physics-based and data-driven parameterisations for nanomaterials and organic molecules. We describe the underlying theoretical background and the package functionality from the design of multi-component NPs through to the evaluation of the corona.  
\end{abstract}

\maketitle

\section{\label{sec:introduction}Introduction}
Advanced materials represent a new paradigm in materials science: substances with highly specific features and enhanced target properties derived from precise control over their structure and composition. A particularly relevant set of examples of these materials are nanomaterials, which may exhibit properties that significantly differ from the expected behaviour of the same bulk material due to the high surface-to-volume ratio. The large surface implies high specific reactivity and capacity for steering complex processes at the molecular level.  
New materials, however, come with new risks: these same desirable properties may also lead to unwanted behaviour when these novel materials come into contact with the environment or living beings \cite{ dobrovolskaia2009,  ilinskaya2013}. As with the benefits, these risks are high for nanomaterials, since their small size enables rapid uptake by the body through multiple pathways, e.g. inhalation, ingestion, or skin contact. Consequently, it is important to be able to predict whether a given material is toxic or biocompatible early at the stage of the material's development \cite{williams2008}. Given the vast range of materials used in modern technology or considered as candidates for applications, and in light of the general need to reduce the amount of in vivo and in vitro tests performed, this suggests the use of in silico methods to predict bioactivity from first principles \cite{lobaskin2023computational}.

To date, the main focus of experimental studies -- and thus the initial goal for computational methods designed to predict experiments -- for the bioactivity of nanoparticles (NP) has been focused on the protein corona: the layer of proteins directly and strongly bound to the surface, the hard corona, of the NP and the soft corona of molecules adsorbed to these inner proteins \cite{kopac2021protein, hajipour2023overview}. Recently, growing attention has also been paid to the fact that the corona need not consist only of proteins \cite{docter2015nanoparticle}. Other molecules, be they metabolites, peptide fragments, lipids, or small organic molecules such as hormones, medicine, or toxins will also adsorb to the NP and likewise be transported along with it, and these may completely alter the final destination and biological outcomes, whether this is deliberate (as in a drug nanocarrier) or accidental. Thus, the computational methodology relating to the corona should be sufficiently general to account for a wide variety of biomolecules, or indeed arbitrary organic molecules, in addition to proteins. The relative binding affinity of these constituents and, hence, the corona composition is controlled by the molecular-level properties of the NP surface: the type of atoms and their connectivity, their partial charge, polarisability, density, and larger-scale geometrical features such as the crystal structure and its curvature. To be able to connect these properties to the corona composition, these must all be factored into the computational methodology for corona prediction. Moreover, in the context of advanced materials, the methodology must allow for the combination of various structural elements (core, shell, dopants, functional groups) defined in terms of their specific properties into a composite NP that reflects its structural complexity and relates it to the characteristics of the corona. 

A key challenge on this path is bridging the length- and timescale gaps between the fundamental features of the materials and the high-level properties of interest, which can amount to several orders of magnitude \cite{lobaskin2023computational}. On the one hand, the adsorption of a single molecule to an NP is highly dependent on the local atomic structure and presence of solvent, and thus this requires the use of atomistic-level methods. The quantum and classical atomistic methods, however, require enormous resources to scale up to the adsorption of a single protein for the timescale of milliseconds. On the other hand, a typical corona may consist of hundreds of adsorbates and develops over the course of hours. Thus, it would be prohibitively expensive to simulate the corona for a single NP in a single medium, let alone scan over multiple NPs or biological environments using a brute-force atomistic simulation. To overcome this, it is necessary to employ coarse-graining methods to allow for longer timescales and larger systems to be reached while maintaining physical accuracy and connection to the original material. Given the wide range of potential adsorbates and NPs, these methods must be sufficiently generic to cover as many possibilities as possible while remaining accessible enough such that a novice user can perform corona predictions without extensive training. A key advantage is granted by the fact that the vast majority of biomolecules, and proteins in particular, can be represented using a relatively small number of simple repeat units such as amino acids (AA) or sugars. It is reasonable to precalculate the interaction for these building blocks and use these to construct a model for the entire biomolecule, or family of related biomolecules, thus greatly reducing the amount of effort that must be expended to evaluate the total adsorption energy or parameterise a new biomolecule. Likewise, although NPs can in principle be highly complex, they too can be subdivided into interchangeable components such as solid or hollow spheres or cylinders of simple materials, and these can be used to construct multicomponent NPs step-by-step. This has led us to the development of a series of increasingly complex models for protein adsorption, starting from an initial simple model using Lennard-Jones (LJ)-like interactions between NP beads and AAs \cite{lopez2015coarse}, to a more complex model of protein adsorption to gold \cite{power2019multiscale} or titanium dioxide \cite{rouse2021} and more recently including multiple NP components simultaneously \cite{subbotina2023silico}. We have further developed models for the prediction of the corona, taking advantage of binding energies computed using the protein-NP models and allowing for a simulation of competitive adsorption in media with a large number of possible adsorbates  \cite{rouse2021hard, hasenkopf2022computational,parinazaluminium2023}.

\begin{figure}[p]
     \centering
     \includegraphics[width=0.85\textwidth]{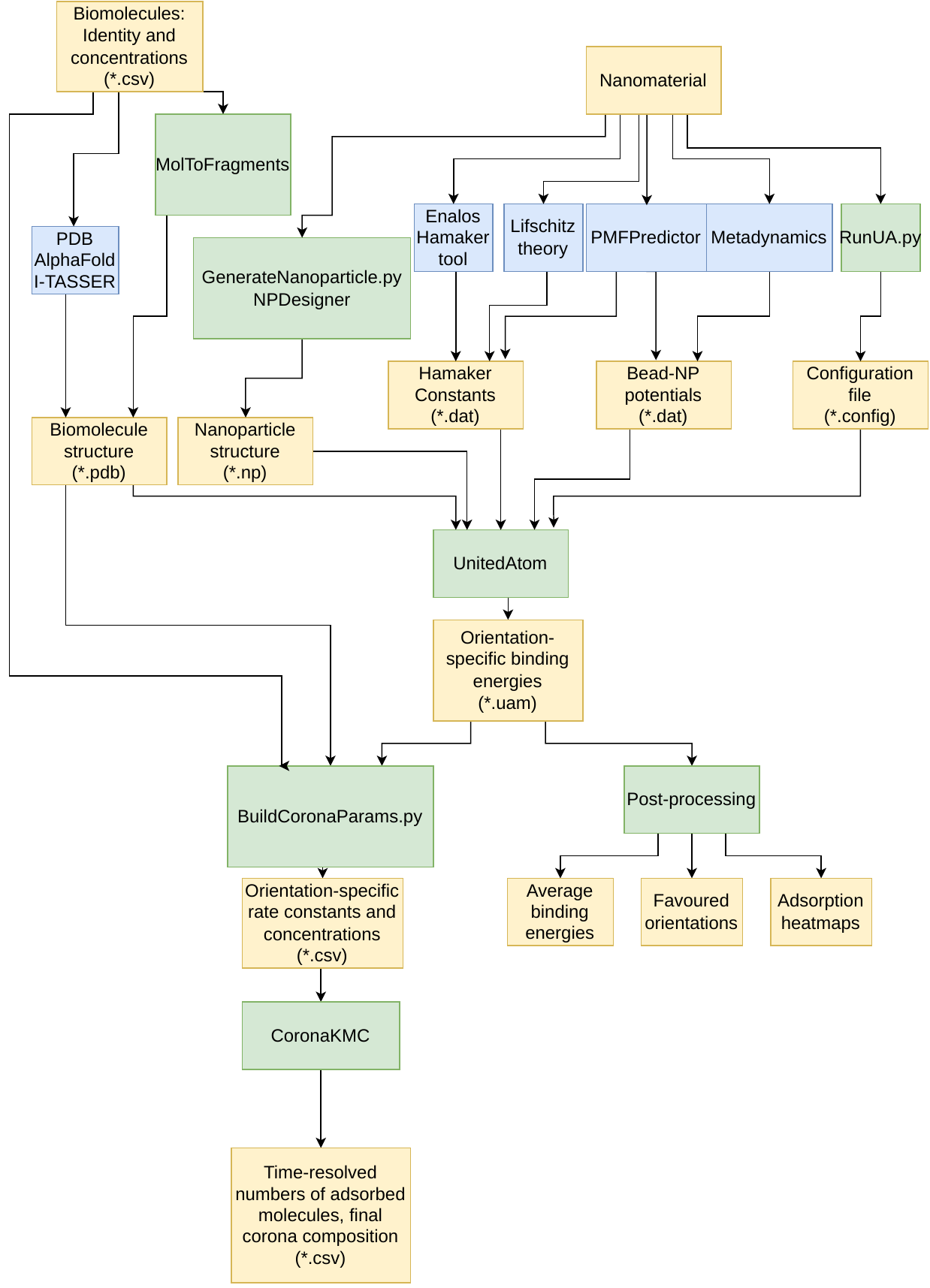}
     \caption{A summary of the overall workflow for corona prediction using the suite of computational tools discussed here. Yellow boxes indicate data used as input/output, green boxes are software tools included in the repository discussed here. Blue boxes indicate other methodologies or software that can be used to produce input as needed.}
     \label{fig:pipeline-summary}
 \end{figure}

In this work, we present a description of \progname{NPCoronaPredict}, the computational pipeline we have developed to enable the prediction of the corona for NPs immersed in solution containing multiple potential adsorbates -- typically, but not necessarily, of biological origin -- with an overview of the generic workflow presented in Figure \ref{fig:pipeline-summary}. The basis of this material-specific prediction is a set of interaction potentials based on the atomistic structure of the NP surface for each material and the small molecules or molecular fragments of interest, taking into account the presence of solvent and ions as necessary. These input potentials are supplied for a range of materials obtained via atomistic simulations and cover the adsorption of AA side chain analogues, lipid fragments and sugars to a range of carbonaceous, metallic, and metal oxide structures \cite{power2019multiscale,rouse2021,saeedimasine2020atomistic,subbotinagold,subbotinasilver,subbotina2023silico}. We stress that the general methodology is by no means limited to these surfaces or biomolecular fragments, or indeed to considering only proteins or other biomolecules. To take advantage of this, the repository also contains a large databank of input potentials generated for a wider range of surfaces and approximately 200 small organic molecules generated via a machine-learning (ML) method (the PMFPredictor toolkit) based on atomistic forcefields for the materials and molecules \cite{rouse2023machine, Rouse_PMFPredictor-Toolkit_2024}. This approach enables the rapid generation of even further input potentials for surfaces given an atomistic forcefield and structure, while new small molecules can be generated and parameterised using the GAFF forcefield \cite{wang2005} and acpype \cite{sousa2012acpype} software via their SMILES code. More generally, the user is free to parameterise their own surfaces or chemicals as required to extend the \progname{NPCoronaPredict} software suite to their own particular needs through their own preferred methodology, and the software is designed to be agnostic to the source of these inputs, although we recommend the use of the PMFPredictor methodology since this is designed to produce output compatible with \progname{NPCoronaPredict} and is available open-source \cite{Rouse_PMFPredictor-Toolkit_2024}. We further extend the functionality by providing software tools for decomposing larger organic molecules, e.g. drug candidates, into fragment-based models compatible with this software.

Our multiscale approach has been developed to take advantage of the \progname{UnitedAtom} and \progname{CoronaKMC} methodologies first described elsewhere while expanding these to cover a far greater range of use cases beyond the adsorption of proteins first considered, and provides a convenient pipeline to enable prediction of the corona with minimal user intervention. In particular, the corona for a wide range of simple NPs consisting of a single material type and fixed radius can be generated for a target mixture of proteins and other biomolecules by running a single command or via a graphical interface. We have also significantly improved the ability of the software to handle complex cases such as proteins with concave or hollow regions into which a small NP may dock. The repository as described in this work can be obtained via git at \cite{NPCoronaPredict-repo} and corresponds to Release v1.0.0-alpha. An archived version is maintained at the former \progname{UnitedAtom} package location \cite{bitbucketUnitedAtom} but is no longer maintained. These versions of the software package are consistent with the descriptions provided here; since the code remains in active development future versions may have altered behaviour. A C++ compiler with the boost libraries and headers installed is required to compile \progname{UnitedAtom}, while \progname{CoronaKMC} requires a \progname{Python 3} installation with full dependencies given in the documentation. A QT installation is also required to compile optional graphical interfaces. 

In the following sections, we provide a detailed description of each component of the suite, including the underlying methodology, required inputs and expected outputs as well as examples of usage and validation of individual components.

\section{\label{sec:ua}UnitedAtom: biomolecule -- nanoparticle adsorption affinity}
The \progname{UnitedAtom} (hereafter UA and stylised as a single word to distinguish from generic united atom methodologies) software tool is designed to efficiently calculate the adsorption energy of rigid biomolecules consisting of well-defined repeat units (coarse-grained (CG) beads) to an NP, itself potentially consisting of multiple components (NP beads). Throughout, we use ``biomolecule'' to refer to a rigid structure consisting of one or more CG beads but note it does not have to be of biological origin. The general methodology has been published in detail elsewhere \cite{power2019multiscale,rouse2021,subbotina2023silico}   and is summarised in this work. In overview, for a specific simulation configuration  and input set consisting of a biomolecule structure, a set of NP beads, the pairwise interaction potentials between CG and NP beads  are summed together and integrated over to produce orientation-specific adsorption energy for the biomolecule to the NP, with the set of binding energies at multiple orientations saved as output. In the following sections, we provide an overview of the methodology and discuss the required inputs in more detail.

\subsection{\label{sec:uamethod}Methodology}
When executed, the UA program performs the following steps: generation of potentials for each type of CG bead as a function of the NP beads, generation of samples of different orientations of the biomolecule relative to the NP, summation of the interaction potentials over all the CG beads as a function of their position in the biomolecule - NP complex, and integration of the resulting total potential over distance to produce the adsorption energy. 

Firstly, the NP structure is generated or loaded. A bounding radius $R^b_0$, representing the solid core of the NP is computed from the outer radius of the largest NP bead if this is not manually set in the configuration file. An  outermost bounding radius $R^b_1$ is computed from $\mathrm{max}(|r_n| +R_n)$, where $|r_n|$ is the distance of the bead centre from the origin and $R_n$ is the radius of the NP bead, if this value has not been manually set. This methodology is chosen to produce reasonable results for an NP consisting of a core and brush configuration, for which the inner radius encapsulates the core and the outer radius ensures that all of the brush is included. We note that these automatically assigned values can be manually overridden if desired, which may be necessary for NPs without a  well-defined core such as agglomerates. Next, the required interaction parameters for all target CG bead types are loaded and the interaction potentials between each type of NP bead and CG bead are computed. In the default methodology, the interaction between a CG bead type (ALA, GLY, etc) indexed $m$ and the total NP complex is then computed along a single axis by summation over all NP beads indexed $n$, that is:
\begin{equation}
    U_{m}(z) = \sum_{n \in NPs} U_{n,m}(r^* )
\end{equation}
where $r^* =  r^* (x_n,y_n,z_n,z,R_{n})$ is the geometry-specific closest approach of a CG bead centre at $(0,0,z)$ to the NP bead of size $R_{n}$ and with centre at $(x_n,y_n,z_n)$, and the input potentials depend on the particular CG bead -- NP bead pair and are described later. If this pre-summation is disabled, then the computed potentials for each NP bead and CG bead type $U_{n,m}$ are instead stored in memory for later use.  

With the NP structure defined, the biomolecules of interest are loaded sequentially and orientational sampling is performed. First, the biomolecule is shifted such that its centre of mass (COM) is defined to be $(0,0,0)$ and rotations are applied to set the biomolecule to the target orientation. This orientation is defined by two angles $\phi,\theta$ and optionally a third angle $\omega$ if provided. The input structure for a given biomolecule is rotated by an angle equal to $-\phi$ around the $z$ axis, followed by a rotation of $180^\circ - \theta$ degrees around the $ y$ axis. Depending on the selected geometry and configuration options, a final rotation of $\omega$ around the $z$ axis may then be applied; this functionality is disabled for basic spheres by default but can be manually enabled for anisotropic NPs and is automatically enabled for cylindrical NPs. Following this rotation, the vector originally defined by $( \cos \phi \sin \theta, \sin \phi \sin \theta, \cos \theta)$ is mapped to $(0,0,-1$), which is normal to and pointing towards the surface of the NP, while the angle $\omega$ produces a rotation around this axis or, equivalently, a rotation of the NP. Note that if the NP is symmetric with respect to rotation around the $z$ axis the rotation around $\omega$  will not change the final output. We further note that in the original frame of reference of the biomolecule as specified in the input .pdb file, the NP is located at spherical coordinates given by $\phi,\theta$. The biomolecule is then translated along the line $(0,0,z)$ to define its location at a fixed NP-offset distance $h$, where a range of values of $h$ are sampled during the calculation according to limits discussed later.

In the default pre-summation model, the NP-Complex -- biomolecule potential is then obtained by summation of $U_m(z)$ potential over all CG beads indexed $i$,
\begin{equation}
    U(h) = \sum_{i \in CGs} \alpha_i U_{m(i)} (  x_i, y_i, z_i + h )
\end{equation}
where  $x_i,y_i,z_i$ are the bead locations defined by the geometry of the molecule and $h,\phi,\theta,\omega$, the CG bead type for bead $i$ is denoted $m(i)$, and $\alpha_i$ is a per-residue weight, with $x_i,y_i,z_i,\alpha_i$ read from the input file as discussed in Section \ref{sec:biomoleculedef}. This default behaviour performs acceptably well for isotropic NPs but does not produce meaningful results for an NP decorated with a brush, for which the potential experienced by a given CG bead depends on all coordinates and not just its distance from the NP.  In this case, we recommend disabling pre-summation such that the potential is instead given by:
\begin{equation}
    U(h) =\sum_{n \in NPs} \sum_{i \in CGs} \alpha_i U_{ n, m(i)} (  x_n,y_n,z_n, x_i, y_i, z_i + h ).
\end{equation}
This double summation substantially increases the required computational time but produces a more physically accurate result. If pre-summation is not disabled, the physical geometry of the NP and presence of brush beads is accounted for in a less accurate way by imposing a small ``overlap penalty'' if any CG bead is determined to be in a location which would overlap with one of the NP beads. For reasons of numerical stability, an additional extreme-short-range potential is also applied in both of these summation models to prevent the calculation diverging when sampling reasons of space in which an NP and CG bead overlap, which we define to be when the centre of the CG bead is at a distance of under $0.1$ nm from the surface of the NP bead. This potential is not directly parameterisable by the user and has the form $U_x(h) = (0.1/h)^{12} - 1.0$ in units $k_BT$ and is set to $0$ for $h>0.1$ nm such that it does not contribute to the potential in realistic conformations.

Once the potential summed over all CG beads is obtained, UA then performs a free-energy integration,
\begin{equation}
    E_{ads} = - \mkbt \ln \left[ \frac{\int_{R_{min}}^{R_{max}} \xi^\alpha e^{-U[h(\chi)]/\mkbt} d \xi}{\int_{R_{min}}^{R_{max}}  \xi^\alpha d \xi} \right]
\end{equation}
where $\alpha = 2$ for spherical coordinates, $\alpha=1$ for cylindrical coordinates and $\alpha = 0$ for planar systems, and we use the variable $\xi$ to represent the distance from the centre of the NP to the COM of the biomolecule. By default, $T = 300$ K but this parameter may be set in the configuration file. The bounds of integration are automatically chosen based on the NP geometry and the structure of the biomolecule as follows. The inner bound $R_{min}$ gives the closest approach of the COM of the biomolecule to the COM of the NP complex and is computed from the structure of the biomolecule and the NP binding radius $R^b_0$ using one of two methodologies. The first, applied by default, is chosen to approximate the situation of the biomolecule approaching from infinity along the $z$-axis and stopping at first contact with the NP, i.e., at the maximum COM-COM distance such that a bead of the biomolecule is in contact with the NP or when the COM-COM distance is equal to zero, whichever occurs first. The second, chosen if the user enables the ``full-scan'' mode, instead chooses this lower bound such that the COM of the biomolecule is placed as close as possible to the COM of the NP without any CG beads existing inside the NP's inner radius. In both cases, the outer bound of integration is then set such that the distance between the plane defined by $(0,0,R^b_1)$ and the lowermost point of the biomolecule is equal to $2$ nm for consistency with previous versions of UnitedAtom and to ensure that all CG beads are an adequate distance from all NP beads that they may be taken to be non-interacting.

This integration is repeated for each target orientation. By default, the angles $\phi,\theta$ are sampled on a grid with $\phi \in [0^\circ,360^\circ]$ and  $\theta \in [0^\circ,180^\circ]$. This grid is divided into units of area $5^\circ \times 5^\circ$, with $64$ points selected at random with uniform density inside each of these units. The adsorption energy is calculated for each of these $64$ sub-samples and averaged together to reduce artefacts and reflect uncertainty in the exact orientation of the protein. By default, this averaging employs a simple, unweighted mean, but the user can optionally enable a mode in which the local energies are averaged by their Boltzmann weights. The resulting average is then reported for the nominal lower limit of the region, that is, the output value for $\phi,\theta$ is the average of $64$ values in the region $[\phi,\phi+5] \times [\theta,\theta+5]$ such that the average value sampled is $\phi+2.5^\circ, \theta+2.5^\circ$, which should be used in post-processing of these results. Note that this oversamples points close to either pole of the sphere with $\theta \rightarrow 0^\circ$ and $\theta \rightarrow 180^\circ$, which must be corrected for when post-processing results as discussed later. For post-processing, we stress the importance of following the correct rotation procedure to avoid misinterpretation of results. Since rotation matrices in three dimensions do not commute and a rotation of $180^\circ - \theta$ produces a very different result to a rotation of $\theta$, it is vital that the rotations are applied in the correct order and using the correct magnitude, $R_z(-\phi)$ followed by $R_y(180^\circ - \theta)$.

\subsection{Configuration file \label{sec:uaconfig}}
\begin{figure}[tb]
     \centering
     \includegraphics[width=0.7\textwidth]{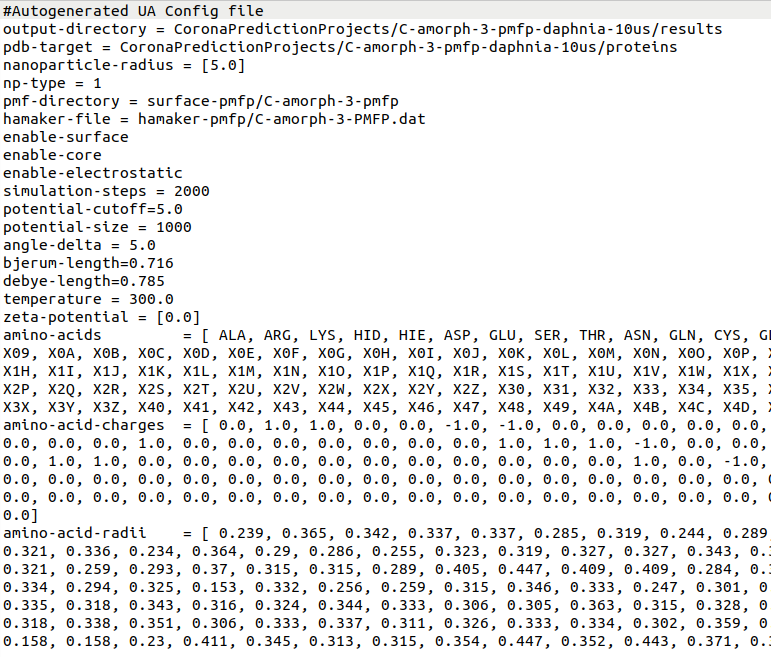}
     \caption{A typical configuration file for a single-component NP \progname{UnitedAtom} run. When executed, this will generate a spherical NP of radius 5 nm using the material parameters for amorphous carbon and compute binding energies for all biomolecules stored in the directory given by pdb-target. }
     \label{fig:config-example}
 \end{figure}

\progname{UnitedAtom} is executed using the command ``UnitedAtom --configuration-file=x.config'', where the configuration file instructs the program where to find all the required inputs and specifies parameters for the calculation, see Table S1 and S2 for further information. This must include either instructions for how to generate single-bead NPs by specifying a range of NP-radii, zeta potentials, and providing the material configuration, or the location of pre-constructed NPs through the use of the np-target option. It must also include via the pdb-target option the location of either a folder containing the set of biomolecule structure files or a single biomolecule. In addition to this, it must contain three lists of values defining the three-letter codes for all CG beads present and their associated charges and radii. The configuration file moreover includes further parameters such as the Debye length, the global NP-shape which defines the choice of the coordinate system (spherical, cylindrical, or planar), and a range of further options to fine-tune simulation settings. An example is shown in Figure \ref{fig:config-example}. The temperature is used only as a scaling factor in the evaluation of adsorption energies; no thermal noise or relaxation is applied to the positions of CG beads and input potentials are not automatically adjusted to account for any thermal effects that may apply there. The configuration file also specifies the location to which all output files should be saved and whether existing files in this directory should be overwritten or the calculation skipped if the output would overwrite a file.

\begin{figure}[tb]
     \centering
     \includegraphics[width=0.7\textwidth]{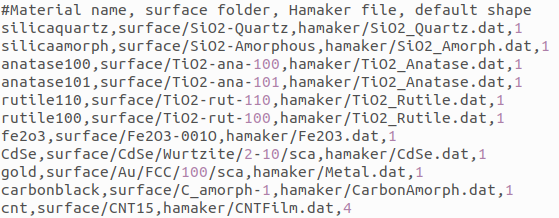}
     \caption{An example of the file defining the types of materials pre-registered with \progname{UnitedAtom}, define by a name, surface potential folder, Hamaker file, and the index for the default shape. }
     \label{fig:materialfile-example}
 \end{figure}

Given the complexity of the configuration file, it is recommended that either a pre-existing template is used,  or the RunUA.py interface is employed to generate a suitable configuration file. This script takes as user input a folder containing target biomolecules and an NP material of interest, together with the radius and zeta potential. We note that the software code interprets the supplied zeta potential as the value of the electrostatic potential at the surface of the NP. This is not always valid and the actual surface electrostatic potential should be used if known. Further options, e.g. the temperature and ionic strength can be set as needed. The available materials are defined in the files  ``MaterialSet.csv'' (see Figure \ref{fig:materialfile-example}), corresponding to materials with input potentials found via metadynamics and ``MaterialSet-PMFP.csv'' for the larger set of materials with potentials obtained via ML with PMFPredictor. The selected material is used to fill in the fields for the pmf-directory, hamaker-file and np-type options to produce a pre-defined material, but this can be edited as needed to allow for custom materials or greater flexibility. If the selected file belongs to MaterialSet.csv then bead parameters are read from beadsets/StandardAABeadSet.csv and used to fill in the corresponding lines in the configuration file, whereas the parameters in pmfp-beadsetdef/PMFP-BeadSet.csv are used otherwise. These latter parameters are more approximate and automatically generated for a much larger variety of beads based on the summation of forcefield parameters and approximate combination rules. At present, the RunUA.py interface is primarily designed for single-bead NPs, but a folder of multicomponent NPs can be passed as an additional input. We further provide a graphical user interface \progname{UAQuickRun} to simplify running UA for simple cases consisting of a single NP and small sets of biomolecules of interest. This interface allows the user to set a few parameters, e.g. a target NP file or material and radius, and a biomolecule PDB file, and passes these as input to the RunUA.py script. The interface further automates the fetching of protein structures from the RSC PDB (if given a four-letter code) or the AlphaFold ensemble database (if given a UniProt ID) and can prepare input for \progname{CoronaKMC} runs. Further options in the interface allow for the plotting of heatmaps from result files and  visualisation of the biomolecule-NP configuration at selected orientations for an initial examination of results. 

 \begin{figure}[tb]
     \centering
     \includegraphics[width=0.5\textwidth]{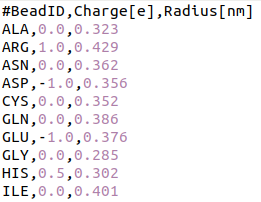}
     \caption{An extract from a CG bead definition file, listing the three-letter code, charge and radius of each CG bead type which can be included in simulations.}
     \label{fig:beadfile-example}
 \end{figure}

\subsection{Biomolecule definition \label{sec:biomoleculedef}}
The biomolecule of interest is represented as a list of atomic coordinates using the PDB file format, with the three-letter residue code used to identify the specific set of interactions to employ for a bead at that location specified by the $x/y/z$ coordinates (in Angstroms) fields. To allow for rapid coarse-graining of proteins obtained from the PDB repository, only lines starting with ``ATOM'' are processed and of these, only atoms labelled as ``CA'' are registered by the program. All other biomolecule files must provide beads following this convention and the same formatting with column widths defined as in the standard PDB specification to ensure these are read correctly. The standard AA residue codes e.g. ALA are reserved to describe AAs specifically within the context of a protein, for which UA by default employs the convention that long-range interactions are treated as if it was a full AA, while short-range interactions are computed using only the side-chain analogue (SCA) under the assumption the amino functional group is fixed within the backbone of the protein and not available for direct binding to the surface of the NP. Two AAs, glycine and proline, require special treatment as they do not have a well-defined SCA. Depending on the source of PMFs, glycine is either omitted or treated as a full AA and proline is either represented as cyclopropane or a full AA. The HIS bead is likewise a special case due to its range of protonation states. In the default bead set it is assigned a charge of $0.5e$ and the PMFs are typically supplied for delta (HID), epsilon (HIE) and fully protonated (HIP) configurations, one of which must be used for the HIS bead unless pre-processing is performed as discussed later.   Other molecules may be represented using any three-letter tag, and a list of suggested assignments is included in the repository for the extended beat set covering a range of common biomolecular fragments. For example, in this set the bead code ``X1B'' corresponds to a CH4 molecule (alanine side-chain) for both PMF and Hamaker constant, while ``X6H'' produces full-AA alanine for both PMF and Hamaker constant.  The occupation field is used to scale the total potential for that particular bead, which may be used to represent disordered residues or molecules that may potentially exhibit multiple charge states at a given pH. It is essential that this column is correctly defined to avoid reading in an erroneous value which is interpreted as a $0.0$, thus leading to that bead being ignored entirely. The b-factor column is used to identify beads which do not occupy a well-defined position in the biomolecule which can be optionally disabled entirely or moved to the centre of the molecule, depending on options set in the UA configuration file. The remaining PDB fields are presently unused.

\begin{figure}[tb]
     \centering
     \includegraphics[width=0.79\textwidth]{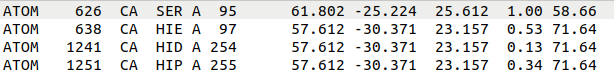}
     \caption{A sample input structure defining a biomolecule, extracted from the PDB repository for 1AX8. Here, a histidine residue has been mapped to three CG beads reflecting different protonation states with weights stored in the occupation column, while a serine bead is provided with a weight of unity as it has no alternate states. Note that most beads have been omitted for clarity.}
     \label{fig:pdbfile-example}
 \end{figure}

Typically, a standard protein structure file obtained from the PDB \cite{berman2000protein}, AlphaFold  \cite{jumper2021highly, varadi2022alphafold}, I-TASSER \cite{yang2015tasser} or most other sources will be directly compatible with \progname{UnitedAtom}, provided it adheres to the standard PDB file format as discussed above with fixed-width columns as provided in the PDB specification.  Optionally, pre-processing can be applied using the included script PreprocessProteins.py. This pre-processing consists of two stages. Firstly, the PROPKA software tool (if available) is run to compute the pKa of residues which may exist in multiple charge states. The target pH of the solution is then employed to predict the probability for the residue to exist in each of these states, and the residue is replaced with multiple ``fractional beads'' according to this weight. For example, a histidine residue (HIS) is replaced by three beads corresponding to the epsilon-histidine (HIE), delta-histidine (HID), and protonated histidine (HIP) beads to avoid the requirement to manually select a suitable HIS potential, see Fig. \ref{fig:pdbfile-example} for an example.  Secondly, a principal axes transformation is performed to rotate the protein into ``canonical form'', in which the $z$-axis  is associated with the smallest moment of inertia and the $y$-axis with the second smallest, and further rotations are performed to ensure that the projection of the electric dipole moment onto these two axes is positive. The goal of this set of rotations is to provide a well-defined initial state for the protein such that later rotations are defined with respect to this rather than the potentially random orientation obtained from the initial source. 

\begin{figure}[tb]
     \centering
     \includegraphics[width=0.7\textwidth]{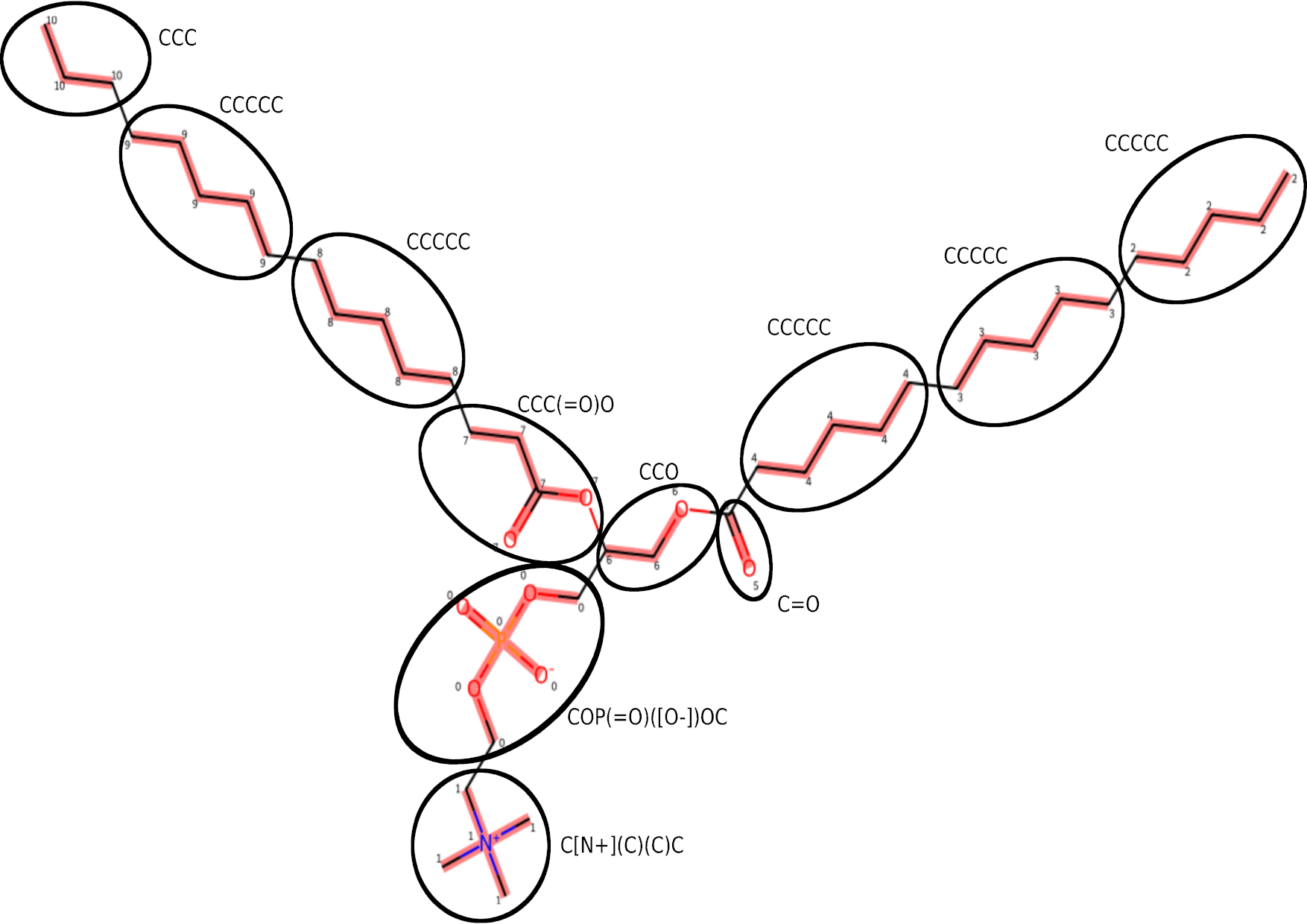}
     \caption{Automatically generated CG bead mapping for a target molecule (DPPC) using the MolToFragment.py script included in the repository, with highlighting applied to indicate the resulting fragments and manual annotation added to indicate SMILES codes for each bead. The mapping has been constrained to use only bead types for which interaction potentials are available.}
     \label{fig:beadmapping-example}
 \end{figure}

For non-protein biomolecules, a suitable CG representation is not necessarily available. Single-bead models for all the small molecules/biomolecular fragments are provided in pmfp-beads.zip, while larger molecules must be represented in terms of these available beads. For these more complex cases, we have developed a script (MolToFragment.py) to produce input compatible with UA based on matching fragments of an input molecule to pre-defined beads, with an example shown in Figure \ref{fig:beadmapping-example} using the ``ForwardsMatching'' algorithm included in this script. In brief, this script attempts to break down a target molecule into smaller fragments by matching SMILES codes of potential fragments to those corresponding to molecules which have already been parameterised. Where possible, we recommend the use of expert knowledge to produce mappings, since this may identify symmetries that the matching algorithm does not identify and this algorithm may require breaking ring structures to achieve a match. For more advanced users, additional modes are implemented for which generated splittings do not need to correspond to pre-existing fragments, e.g. the  ``EqualParts'' method in MolToFragment.py or the BRICs method as implemented in rdkit \cite{rdkit}. These will typically require the production of new interaction parameters but may produce more physically meaningful representations as these methods will not break rings. For manual generation of biomolecules, it is highly recommended to adapt existing templates to ensure that all fields are located in the correct columns; in particular, if the occupancy field is misaligned the bead will typically be assigned an occupancy of zero and thus not contribute to the binding energy.

\subsection{NP definition \label{sec:uanp}}
\begin{figure}[tb]
     \centering
     \includegraphics[width=0.7\textwidth]{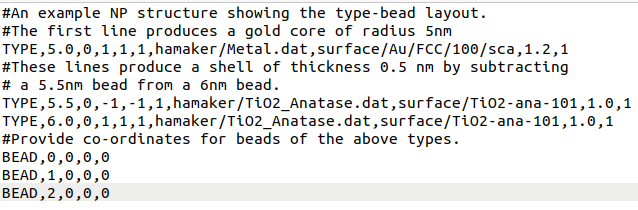}
     \caption{A sample NP definition file to produce a gold core (bead type 0) with a thin layer of anatase, generated by subtraction of a bead of radius $5.5$ nm from a bead of radius $6$ nm.}
     \label{fig:npfile-example}
 \end{figure}

\begin{figure}[tb]
     \centering
     \includegraphics[width=0.89\textwidth]{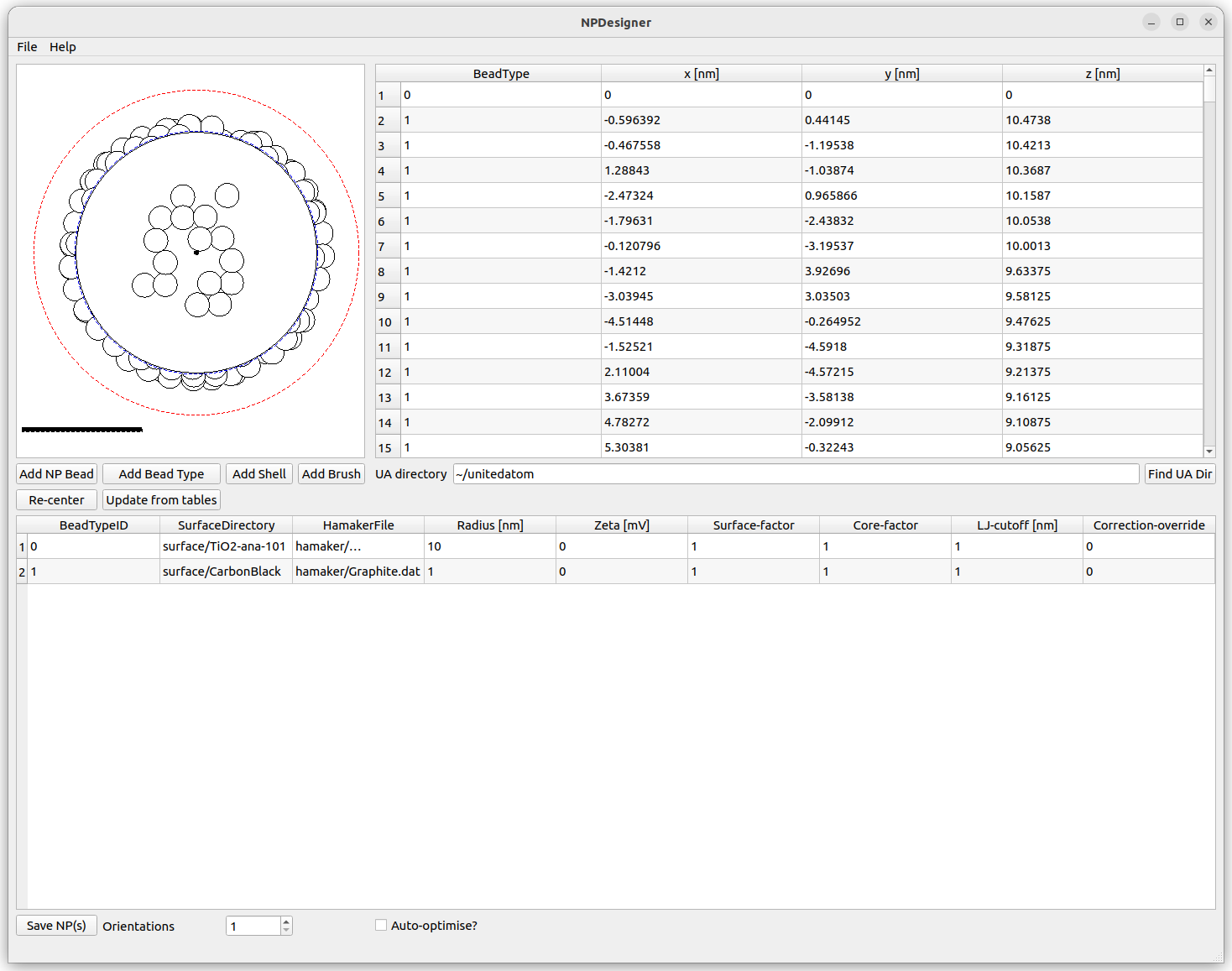}
     \caption{An example NP consisting of an anatase core decorated with carbon black beads produced using the NPDesigner software tool. The locations of beads are shown in the right-hand table, while the bottom table lists definitions of all bead types which have been added so far. A visualisation of the NP is shown in the upper left corner, with the dashed blue line indicating the NP bounding radius at the nominal surface and the red dashed line indicating the limit at which adsorbates are assumed to be unbound. }
     \label{fig:npdesigner-example}
 \end{figure}

Simple NPs consisting of a single component can be defined directly in the configuration file using the ``radius'' and ``zeta-potential'' lists together with a specified Hamaker file and surface potential directory, with UA automatically generating all combinations of these two for the material and shape in question. This further requires setting the np-shape parameter, which takes the value $1$ to produce a sphere, $2$ for a solid cylinder with planar-to-cylindrical potential mapping (see next section), $3$ for a cube (planar mapping), $4$ for tube  (cylinder-to-tube mapping) and $5$ for a solid cylinder (cylinder-to-cylinder mapping). Here, a tube is a hollow cylinder suitable as a model for single-wall carbon nanotubes (CNTs) while cylinders have a solid centre to represent elongated NPs or multi-wall CNTs. For more advanced NPs consisting of multiple components, e.g. a core with a shell or a brush, or an agglomeration of smaller NPs, the NP is defined using a specialised file format to instruct UA on the location and nature of all NP-beads, with a simple example shown in Figure \ref{fig:npfile-example}, and the np-shape option sets the global coordinate system and it is generally recommended that this be set to the spherical value. These files can be constructed manually and descriptions of the required file format are provided in the documentation in the repository, or the supplied GenerateNanoparticle.py script can be employed to generate NPs according to pre-defined combinations of shells and brush densities. A graphical tool NPDesigner (Fig. \ref{fig:npdesigner-example}) is also provided to simplify the production of common NP configurations, i.e., combinations of single beads, shells and brushes, with brushes generated one layer at a time with beads placed at locations using the algorithm presented in \cite{koay2011analytically}, with the output produced in either the .np format required for UA or in .pdb format for ease of visualisation. Note that the NP is not rotated in UA itself, except for an effective rotation applied by the rotation of the biomolecule by an angle $\omega$ if this is enabled, which is equivalent to rotating the NP around the $z$ axis by $-\omega$. If more complex orientations are required they must be supplied as extra .np files with the rotation applied manually, using e.g. Arvo's algorithm to produce rotations which result in an isotropic distribution of new orientations \cite{arvorotation}. This algorithm is implemented in NPDesigner to allow for the production of multiple output files corresponding to the same NP in different orientations.

\subsection{Input potentials \label{sec:uapot}}
\begin{figure}[tb]
     \centering
     \includegraphics[width=0.7\textwidth]{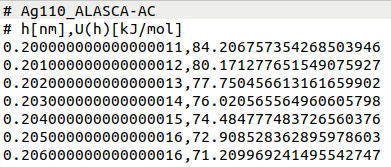}
     \caption{An example of the input required to describe the short-range surface potential for a particular CG bead (here an alanine side chain analogue) to a given NP type (silver, 110 surface plane). }
     \label{fig:pmffile-example}
 \end{figure}

UA requires parameters for the interaction potentials for each biomolecule bead type with each NP bead type. Three main classes of potentials are used in the UA framework, with the potential for a CG bead of type $i$ with an NP bead of type $n$ given by,
 \begin{equation}
     U_{i,n}(d) = U_{S,i,n}(d) + U_{H,i,n}(d) + U_{el,i,n}(d)
 \end{equation}
where $d$ is the distance of closest approach between the beads, $U_{S}$ is a tabulated short-range (surface) potential corresponding to the interaction between the CG bead and the surface of the NP, $U_{H}$ is a Hamaker-like (integrated vdW) potential and $U_{el}$ is an electrostatic potential.  The tabulated short-range potential $U_S$ must be provided for each NP material and CG bead by specifying a folder for that material containing a set of files XXX.dat, where XXX is the three-letter code associated with that CG bead and must be consistent with the definition used in the configuration file and biomolecule structure file to ensure that UA assigns the correct potential to each bead. Each PMF file should contain a comma-separated table of values for the potential (units \kjmol) as a function of the distance of the centre of the bead to the surface of the NP (units nm) as shown in Figure \ref{fig:pmffile-example}. We note that some older PMFs use a fixed-width file format, which remains functional within UA for backwards compatibility but should be considered deprecated in favour of comma-separated files.

This file is typically assumed to be a PMF obtained via metadynamics or ML methods, but can in principle correspond to any distance-dependent potential to apply. Since, in general, these potentials are computed for a particular NP topography (e.g. a planar surface or a cylinder of predetermined radius), a correction function is applied to map these to the expected potential generated by the actual NP of interest, e.g., mapping from a planar configuration to a spherical NP of the given radius. This correction function is generated under the approximation that the main contribution to the surface potential arises from the $1/r^6$ vdW term, such that the ratio of the tabulated potential generated by a volume $v_1$ to that generated by the volume $v_2$ is approximately equal to the ratio of the $r^{-6}$ potential integrated over these two volumes,
\begin{equation}
\frac{U_{S}( r, v_1 )}{U_{S}(r,v_2)} = \frac{ \int_{v_1} (r-r_1)^{-6} d r_1  }{\int_{v_2} (r-r_2)^{-6} d r_2}
\end{equation}
where the integration runs over all points $r_i$ in the volume region $v_i$ \cite{hamaker1937london,israelachvili2011intermolecular,power2019multiscale}.  This correction is implemented for the plane-to-sphere geometry as originally discussed in \cite{power2019multiscale} and has been extended to plane-to-cylinder, cylinder-to-tube, and cylinder-to-cylinder geometries. Alternatively, no correction can be applied, which is required for PMFs generated for small polymer beads. For PMFs generated for other geometries, it is recommended to manually map these to a planar configuration such that UA can automatically re-map them to the target configuration as required.  

\begin{figure}[tb]
     \centering
     \includegraphics[width=0.6\textwidth]{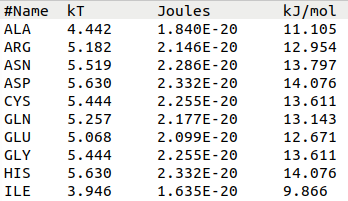}
     \caption{An extract from a file used to parameterise the long-range NP-CG bead potentials, listing Hamaker constants for a range of biomolecular fragments to SiO$_2$ quartz across an aqueous medium. }
     \label{fig:hamakerfile-example}
 \end{figure}

The tabulated short-range potential is assumed to correspond to only a fraction of the total volume of the NP close to the CG beads. To account for the rest of the NP, UA generates a long-range Hamaker-like potential $U_H$ corresponding to the integration of the vdW potential over the volume of the NP and CG bead \cite{hamaker1937london,israelachvili2011intermolecular}. Unlike the traditional Hamaker approach, the integration is performed only over elements of each bead separated by a distance greater than a cutoff distance $r_c$ and is not limited to sphere-sphere interactions only, with the generic integral given by,
\begin{equation} \label{eq:hamakerxgeneric}
    U_{H} = \frac{A_H}{\pi^2} \int_{V_{CG}} \int_{V_{NP}} \frac{ \Theta_h(|r_{NP} - r_{CG}| - r_c) }{\left( |r_{NP} - r_{CG}| \right)^{6}}  d r_{NP} d r_{CG},
\end{equation}
where $A_H$ is the Hamaker constant for that particular CG-NP pair interacting through water \cite{hamaker1937london}, $\Theta_h (x)$ is the Heaviside theta function, used to set the integral to zero within the exclusion region, $r_c$ is the cutoff distance at which the interaction is covered in the PMF (assumed to be equal to the LJ cutoff in the metadynamics simulation) $r_{NP}$ is a point in the NP, $r_{CG}$ a point in the CG bead, and the integration runs over all pairs of points. This produces a function that smoothly switches between different regimes as necessary without a discontinuity at distances of $r_c$ which would be introduced if the interaction is simply switched on once the bead centre is sufficiently far from the NP. For a spherical NP at long range Eq. \eqref{eq:hamakerxgeneric} reduces to the standard Hamaker expression, whereas different results are obtained for cylindrical geometries or for configurations at close range to avoid double-counting elements of the CG or NP beads. For cylindrical geometries, this expression is evaluated partially numerically due to the lack of a closed-form analytical result. The required Hamaker constants are supplied in an input Hamaker file for each NP material, consisting of one line per CG bead type (again requiring consistent three-letter codes to enable correct identification of beads), with the required input format specified in src/HamakerFile.h as depicted in Figure \ref{fig:hamakerfile-example}. These can be computed through Lifshitz theory \cite{israelachvili2011intermolecular} or through summation of forcefield parameters as implemented in either the Enalos Hamaker tool \cite{Afantitis20206523, Varsou2020789} or the scripts supplied in \cite{Rouse_PMFPredictor-Toolkit_2024}, with the latter used to produce Hamaker constants matching the materials with ML surface potentials included in the repository. This also requires the radius for the bead as set in the configuration file, where the radius is typically calculated from forcefield parameters or experimental data as discussed later. 

The third contribution is an electrostatic potential, for which we employ the Debye-H\"{u}ckel approximation to the Poisson-Boltzmann equation to represent the effects of electrostatic screening while allowing for simple analytical expressions for all geometries to be determined. The resulting potentials are defined by the Debye length $\kappa^{-1}$ as specified in the configuration file, the charge of the CG  bead $q_i$, the surface potential $\psi_0$, and the shape of each NP bead. For a spherical NP bead, the resulting potential is given by, 
\begin{equation}
    U_{el} = q_i \psi_0 \frac{R}{R+d} e^{-\kappa d}
\end{equation}
cylindrical by,
\begin{equation}
    U_{el} = q_i \psi_0 \frac{\mathrm{K}_0( \kappa(d + R) )}{\mathrm{K}_0(\kappa R)}
\end{equation}
and planar by,
\begin{equation}
    U_{el} = q_i \psi_0 e^{-\kappa d} ,
\end{equation}
noting that for all three we model the CG bead as a point particle such that $d$ is the distance from the surface of the NP bead to the centre of the CG bead. An expression for finite cubes based on an expansion in terms of spherical harmonics is implemented in the code but is employed only when the size of the NP is on the same order of magnitude as the Debye length, which is generally less than $1$ nm and so the planar potential is typically acceptable. For historical reasons, the value of the surface electrostatic potential is referred to as the zeta potential and a Bjerrum-length parameter is also read in from the configuration file. The Bjerrum length is unused except in certain special cases discussed further in the UA documentation; in typical operation, this parameter can simply be left at its default value as it does not enter into the electrostatic calculations. We also note that many PMFs for charged surfaces already include the effects of the charge-charge interaction and, since the Debye length in UA is typically on the order of $1$ nm, the majority of the surface charge is already accounted for by the PMF. Thus, in some cases, it may be more accurate to set the electrostatic surface potential to $0$ mV to avoid double-counting the charge-charge interactions, unless it is known that the PMFs did not include a charge, e.g. the set of zero-valent metal PMFs or if the electrostatic potential is used to off-set the charge interaction already factored into the PMF to produce a different overall surface charge.

\subsection{\label{sec:output}Output}
\begin{figure}[tb]
     \centering
     \includegraphics[width=0.7\textwidth]{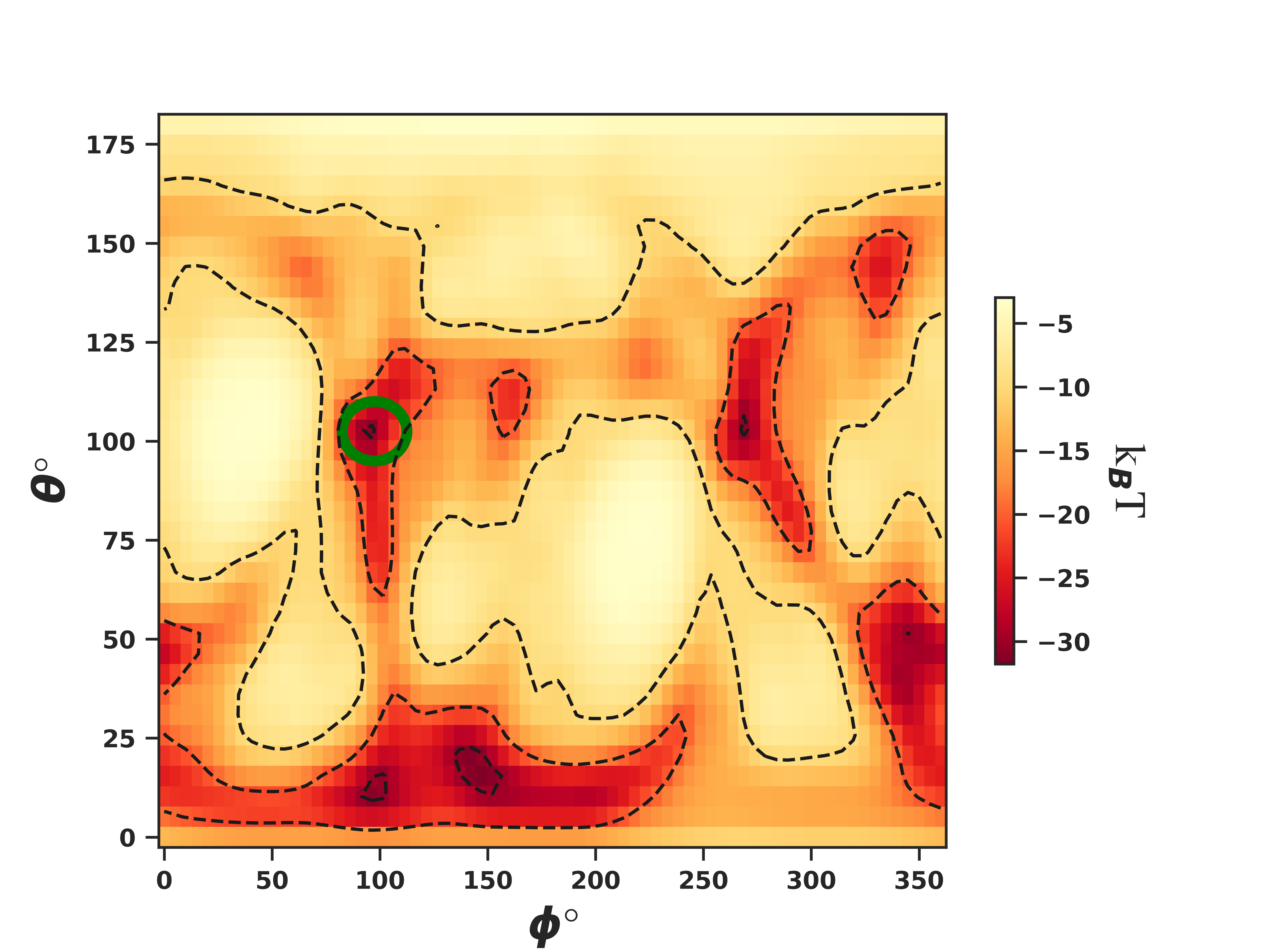}
     \caption{An example heatmap plot of binding energies produced for bovine serum albumin (PDB code 3V03) to a silver NP. The location of the most favourable protein orientation is marked with a green ring at $\phi = 97.5^\circ, \theta=102.5^\circ$}
     \label{fig:heatmap-example}
 \end{figure}

The main output from a UA run is a datafile with an automatically generated filename of the form ``biomolecule\textunderscore radius\textunderscore zeta.uam'' for each NP-biomolecule pair, stored in the designated output folder defined in the configuration file. This datafile contains a table of values mapping each orientational sampling range (given as left-hand edges for $\phi, \theta$ and the fixed value for $\omega$) to a local average of the adsorption energy (provided in units \kbt and \kjmol), the standard deviation of adsorption energies in this interval, mean-first-passage-times (if enabled, else this field contains the value -1), the distance between the nominal surface of the NP and the centre of the closest CG bead, and the average number of residues in close-contact (at a surface to CG bead centre distance of under $0.5$ nm) in that range of orientations. Typically, this datafile is post-processed to provide further results. A very common application is the generation of a heatmap plot as in Figure \ref{fig:heatmap-example} (generated using the script provided in tools/VisualiseUAResults.ipynb ) to highlight the general adsorption affinity of the biomolecule to the NP and identify strongly adsorbing orientations. 

To provide an immediate assessment of the affinity of a given biomolecule to an NP, the binding energy is averaged over orientations according to a given weighting scheme, e.g. the simple average:
\begin{equation}
\langle E \rangle =  \frac{\sum_i \sin \theta_i   E_i}{\sum_i    \sin \theta_i} , 
\end{equation}
or Boltzmann-weighted average,
\begin{equation}
\langle E \rangle = \frac{\sum_i \sin \theta_i e^{-E_i/k_BT} E_i}{\sum_i \sin \theta_i e^{-E_i/k_BT}  }.
\end{equation}
Note that UA output files contain the left-hand edge for $\theta$ and thus these must be offset by $2.5^\circ$ to obtain the central bin, then converted to radians before calculating $\sin \theta$. Of these two averages, the simple average can be thought of as the affinity of an adsorbate which is at a random orientation with respect to the surface of the NP, i.e. during the initial stage of the corona formation. The Boltzmann average, meanwhile, is more strongly weighted towards orientations with high binding affinity, and so reflects the thermal equilibrium achieved in the later stages of corona formation. In certain cases, these averages must be computed taking into account the fact the protein can bind to multiple different surface types, e.g. different crystal facets, Janus particles, or if multiple values of $\omega$ have been sampled for, e.g., CNTs. This is achieved by generalising the above expressions to include an additional weighting term in the numerator and denominator, $w_j$, to reflect the abundance of that particular surface,
\begin{equation}
\langle E \rangle = \frac{\sum_i \sum_j w_j \sin \theta_i e^{-E_i} E_i}{\sum_i \sum_j w_j   \sin \theta_i e^{-E_i} },
\end{equation}
as demonstrated in the MultiSurfaceAverage.py script included in the repository.

\subsection{Pre- and postprocessing}
Here we provide an overview of some additional scripts (Jupyter Notebooks and Python) to provide additional post-processing or visualisation of results obtained using \progname{UnitedAtom}:
\begin{enumerate}
\item \progname{CalcLifschitzHamaker.ipynb} -- an example script to demonstrate the generation of files containing Hamaker constants from optical constants, and bead radii from the coordinates obtained from all-atom models together with the atomic Bondi radii \cite{bondi1964van, zhao2003fast}. 
\item  \progname{VisualiseUAResults.ipynb } -- a script for post-processing of .uam data files to generate and visualise the lowest energy complexes of proteins adsorbed to the surface of the NP. The script produces the PDB file of the complex for further studies, with the protein represented in all-atom resolution and the NP as the component beads, one ``atom'' per bead. An example of the produced visualisation for BSA to a pristine silver NP based on the results shown in Figure \ref{fig:heatmap-example} is shown in Figure \ref{fig:bsa-ag-complex-example}.

\begin{figure}[tb]
     \centering
     \includegraphics[width=0.70\textwidth]{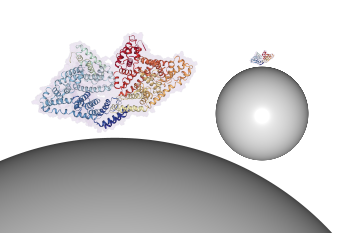}
     \caption{An example of the protein-NP complex produced by post-processing the results of a UA calculation for bovine serum albumin to a silver NP using the \progname{VisualiseUAResults.ipynb } script. The conformation shown is the energetically most favourable orientation of the protein corresponding to the region circled in Figure \ref{fig:heatmap-example}. The inset shows the entire complex while the main figure provides a cropped region to show finer details of the protein. }
     \label{fig:bsa-ag-complex-example}
\end{figure}

\item  \progname{ViewNP.ipynb } -- a script for converting .np files into PDB files, followed by their visualisation. The script generates $*$.pdb and $*$.png files for the input  $*$.np file describing the structure of the NP, with an example for the core-shell PEGylated AgNP previously employed in Ref. \cite{subbotina2023silico} shown in Figure \ref{fig:core-shell-np-example}.

\begin{figure}[tb]
     \centering
     \includegraphics[width=0.35\textwidth]{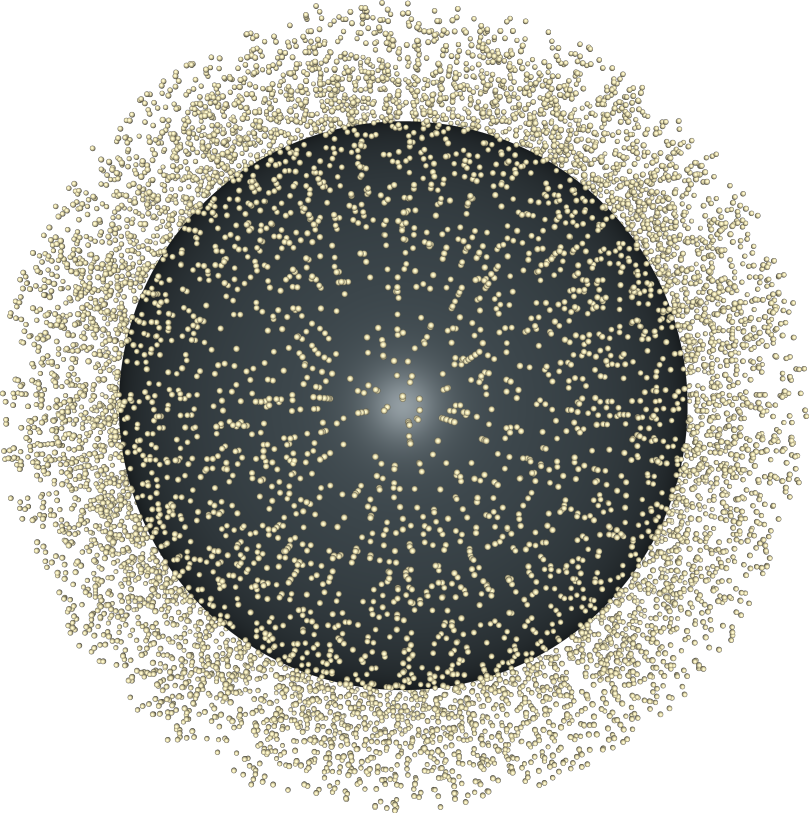}
     \caption{An example of the visualisation for a multi-component core-shell-brush PEGylated silver NP  produced by post-processing of an input  $*$.np file with the  \progname{ViewNP.ipynb } script, with the dark central bead corresponding to the silver core and the small light-coloured beads representing PEG beads. The initial configuration of the NP was obtained by using \progname{GenerateNP.py}.}
     \label{fig:core-shell-np-example}
 \end{figure}

\item \progname{GenerateNanoparticle.py} -- a set of routines for automated production of core-shell-brush type NPs based on pre-defined bead types and brush densities, including raspberry models for inner cores. This tool is recommended for the bulk generation of NPs when it is not practical to individually generate these using NPDesigner. 
\item \progname{ApplyOptimumRotation.py} -- a command line tool for rotation of biomolecules to their optimum binding configuration based on a UA output file, takes command line arguments for batch processing.
\end{enumerate}

\subsection{The UAQuickRun graphical interface}
\begin{figure}[tb]
     \centering
     \includegraphics[width=0.95\textwidth]{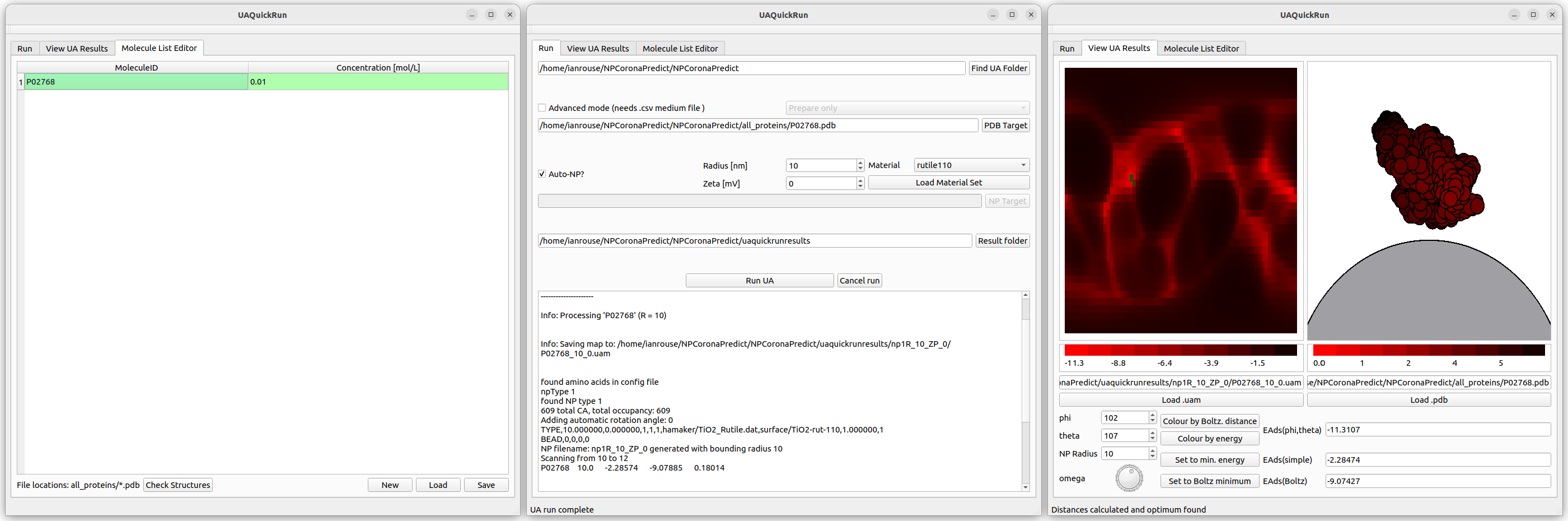}
     \caption{A demonstration of the \progname{UAQuickRun} interface for performing NP-protein binding energy calculations using a simplified set of options. The first panel shows the interface for automatically downloading protein structures based on their ID.  The second one shows the setup and output of computation for a single protein - NP pair (here human serum albumin to a rutile NP), while the third shows the results visualised as a heatmap and schematic view of the favoured orientation. }
     \label{fig:uaquickrun}
 \end{figure}

To assist new users and to allow for rapid interpretation of results, we have designed a simplified GUI named \progname{UAQuickRun} to streamline some of the more common uses for UA as shown in Figure \ref{fig:uaquickrun}. This GUI combines three main tools to simplify the potentially complex procedure. Firstly, a list of biomolecules of interest can be edited on the ``Molecule List Editor'' tab and structures for these found if needed. All structures here are assumed to be proteins, with structures retrieved if needed either from the RCSB PDB \cite{berman2000protein} if they are given a four-character ID code, or from the AlphaFold database \cite{jumper2021highly, varadi2022alphafold} for other protein identifiers. Note that structures can be manually provided if needed and will only be fetched remotely if none is located. The main functionality is located on the ``Run'' tab, which allows the user to generate basic NPs or select an output from NPDesigner, and select a protein (or list of biomolecules) of interest. The GUI can then be used to call the \progname{UnitedAtom} executable and show the results as these are computed. Once a run is complete, the generated data can be visualised as a heatmap and the location of the protein relative to the NP is visualised. We stress that by design, this interface does not incorporate the full functionality of UA, but provides a simplified experience for new users or those less familiar with command-line operations. 

\section{\label{sec:ckmc}CoronaKMC: Corona prediction via kinetic Monte Carlo}
The adsorption affinity of a biomolecule to an NP is not necessarily predictive of its abundance in the corona, especially when there is competition between multiple adsorbing species or orientations of the same species. A large protein may adsorb very strongly but exist in such vanishingly low concentrations compared to other potential adsorbates that its overall abundance remains low, or it may be out-competed by biomolecules that individually adsorb less strongly but occupy a smaller area such that the total energy is more favourable by adsorbing a large number of these, or even be out-competed by another absorbate which binds even more strongly. If, however, no other adsorbates are present then this large protein will then be a major component of the corona. Consequently, a prediction of the corona content must take into account this competition between all adsorbates present. 

A very simple first-order prediction of the corona content may be obtained using the mean-field approximation \cite{dell2010modeling,sahneh2013dynamics}. Given a set of adsorbates $i$ with adsorption free energies $E_i$, concentrations $c_i$, and $n_i$ available binding sites on the surface of the NP, where $n_i$ is inversely proportionate to the cross-sectional area of the adsorbate, the number abundances are approximated by,
\begin{equation}
    N_i = n_i \frac{ c_i e^{-E_i/\mkbt}}{1 + \sum_j c_j e^{-E_j/\mkbt}} .
\end{equation}
This simple expression neglects a number of factors, chiefly, it allows for completely efficient packing of adsorbates onto the surface of the NP and assumes adsorbates can deform to an arbitrary degree. We have previously demonstrated a hard-sphere model of corona formation which overcomes these limitations \cite{rouse2021hard, parinazaluminium2023}. In this section, we discuss the implementation of the kinetic Monte Carlo (KMC) method for evaluating the corona formation as integrated into this package. In brief, this script simulates the sequential adsorption and desorption of adsorbates to the surface of an NP, taking into account factors such as the bulk concentration and availability of free surface area on the NP for binding to take place. The NP is assumed to be a single bead of either spherical, cylindrical or planar geometry with adsorption occurring isotropically across its surface. Thus, if an NP consists of multiple surfaces such as a Wulff structure or a Janus particle, we recommend that a separate simulation is run for each surface type of interest and the total numbers of adsorbed proteins calculated as a weighted sum over all surface types. 

\subsection{\label{sec:kmcinput}Input}
\begin{figure}[tb]
     \centering
     \includegraphics[width=0.9\textwidth]{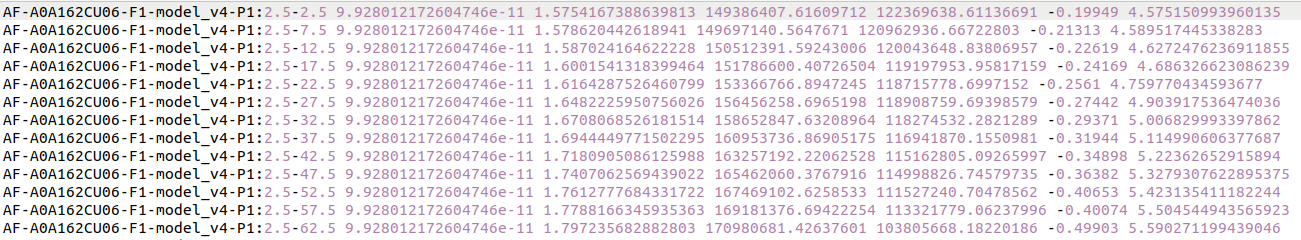}
     \caption{An extract from an input file for the \progname{CoronaKMC} script. The first entry gives a unique identifier for the adsorbate, including its orientation, followed by a concentration in the medium, effective radius, adsorption and desorption rate constants, adsorption affinity, and occupied area. }
     \label{fig:coronakmcinput-example}
 \end{figure}

The most important input to a \progname{CoronaKMC} run is a list of all potential adsorbates, defining their effective size, concentration in the bulk, and rate constants for adsorption and desorption. In simple cases, this file can be manually constructed. In general, however, the BuildCoronaParams script should be employed to automate the conversion of .uam output and .pdb structures to the required input format. This script takes as input a list of biomolecules and their number concentrations in units mol/L, finds matching structures and .uam binding energy tables and computes rate constants and adsorption areas for each orientation of the biomolecule \cite{hasenkopf2022computational}. The output is saved in the structure shown in Fig \ref{fig:coronakmcinput-example}, in which each orientation of a given biomolecule is assigned an individual identification and set of rate constants.

Once an input set has been generated, the script is run using Python3. In addition to the list of adsorbates, further options can be specified as command-line arguments when running \progname{CoronaKMC} to further control the simulation parameters. These options are described in more detail in the documentation and typically enable features such as manual control over the boundary conditions and coordinate system, specification of the amount of simulated time for which the program should run, the use of an algorithm to accelerate the simulation by identification of quasi-equilibrated processes \cite{dybeck2017generalized}, the activation of ``displacement mode'' instead of ``standard mode'' and other parameters.  During the simulation, events are generated corresponding to the adsorption or desorption of adsorbates. Adsorption events correspond to the selection of a potential adsorbate with a probability proportional to the rate at which it collides with the NP and the generation of a random position on the surface of the NP. If this region of the NP is free from adsorbates the particle is accepted automatically in standard mode and with a probability $e^{-E_{ads} / \mkbt }/(1 +e^{-E_{ads}/\mkbt})  $ in displacement mode to reflect the probability that the adsorbate can successfully displace water. If the area of the NP is not free, the adsorbate is rejected in standard mode. In displacement mode, it is accepted with a probability of $e^{-\Delta E / \mkbt }/(1 +e^{-\Delta E/\mkbt})  $, where $\Delta E$ is the difference between the binding energy of the incoming adsorbate and the sum of the binding energies for all currently adsorbed particles which would overlap with the new adsorbate and which must be removed to accept the new adsorbate. It is assumed that water adsorbs to the NP with a reference binding energy of $0 \mkjmol$ and that other adsorbate energies are defined with respect to this. Note that the acceptance probability above is defined such that adsorption which does not change the overall energy is allowed $50\%$ of the time such that both outcomes occur with equal probability. Desorption occurs with a probability dependent on the desorption rate constant; this value is scaled slightly in displacement mode to ensure that the ratio $k_a/k_d$ remains fixed due to the decrease in adsorption for weakly adsorbing biomolecules.

\subsection{\label{sec:kmcoutput}Output}
\begin{figure}[tb]
     \centering
     \includegraphics[width=0.6\textwidth]{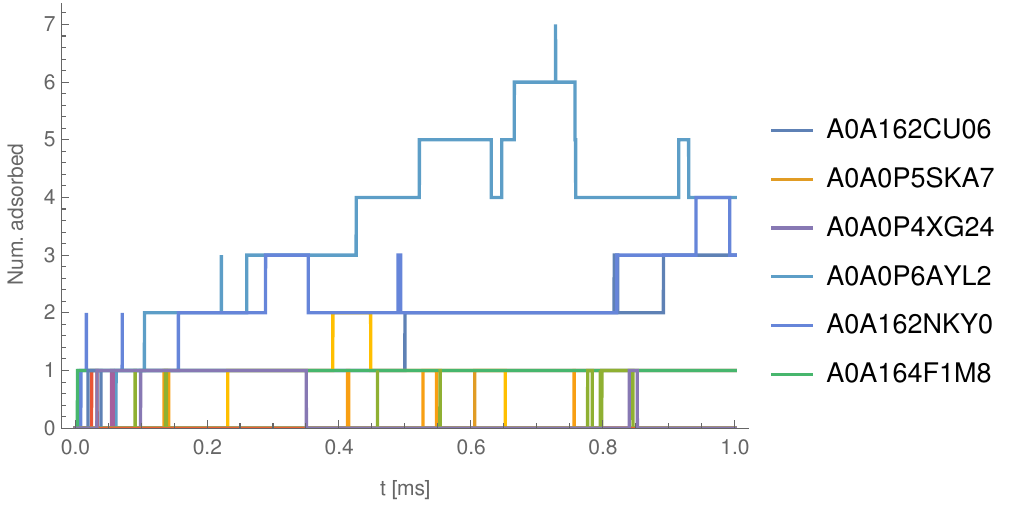}
     \caption{Time-evolution of the predicted corona for a set of twenty proteins (Table \ref{tab:daphniaProteins}) on a 5 nm gold (100) surface. For clarity, only proteins still in the corona at $t=1$ ms have an entry shown in the legend. }
     \label{fig:coronakmc_output}
 \end{figure}

During the runtime of a simulation, the number of each class of adsorbate (summed over different orientations for the same species) is displayed on-screen at predefined intervals, with the same data saved to text files for further use. At the end of a simulation, coordinates for the final corona composition are saved out including the exact identity of each adsorbate to allow for identification of orientations which are present in the corona and for visualisation purposes if necessary. Two such files are generated: one with a .kmc extension, which contains the data in the internal coordinate system and the adsorbate rate constants, and a plain text .txt file with ``finalcoords'' in the filename, which contains the adsorbate name and Cartesian coordinates. An example script tools/CoronaKMCtoVMD.py is provided to convert an output .kmc file for a spherical NP to a .tcl script which can be run in VMD to produce a simple visualisation of the corona as shown in Fig. \ref{fig:coronakmc_output}. More advanced visualisation can be achieved using the BuildCoronaCoords.py script, which produces an output .pdb file containing coordinates based on the atomistic coordinates for all molecules found in the corona in their correct orientations and locations. 

\section{\label{sec:wrapper}NPCoronaPredict: End to end prediction}
In many cases, the same set of biomolecules must be tested against a variety of NPs under essentially identical conditions. To facilitate this, we have developed a wrapper script \progname{NPCoronaPredict.py} (formerly \progname{PrepareKMCInput.py}) to automate running the UA to BuildCoronaParams to \progname{CoronaKMC} pipeline for simple NPs. This script takes as input the list of biomolecules of interest together with their concentrations and automates setting up and performing each step of the calculations, including fetching protein structures from the AlphaFold repository if possible for adsorbate names which correspond to a valid UniProtID. As additional input, it takes the NP size, shape, and material, along with any other parameters to pass to UA or \progname{CoronaKMC} as necessary. 

\begin{figure}[tb]
     \centering
     \includegraphics[width=0.6\textwidth]{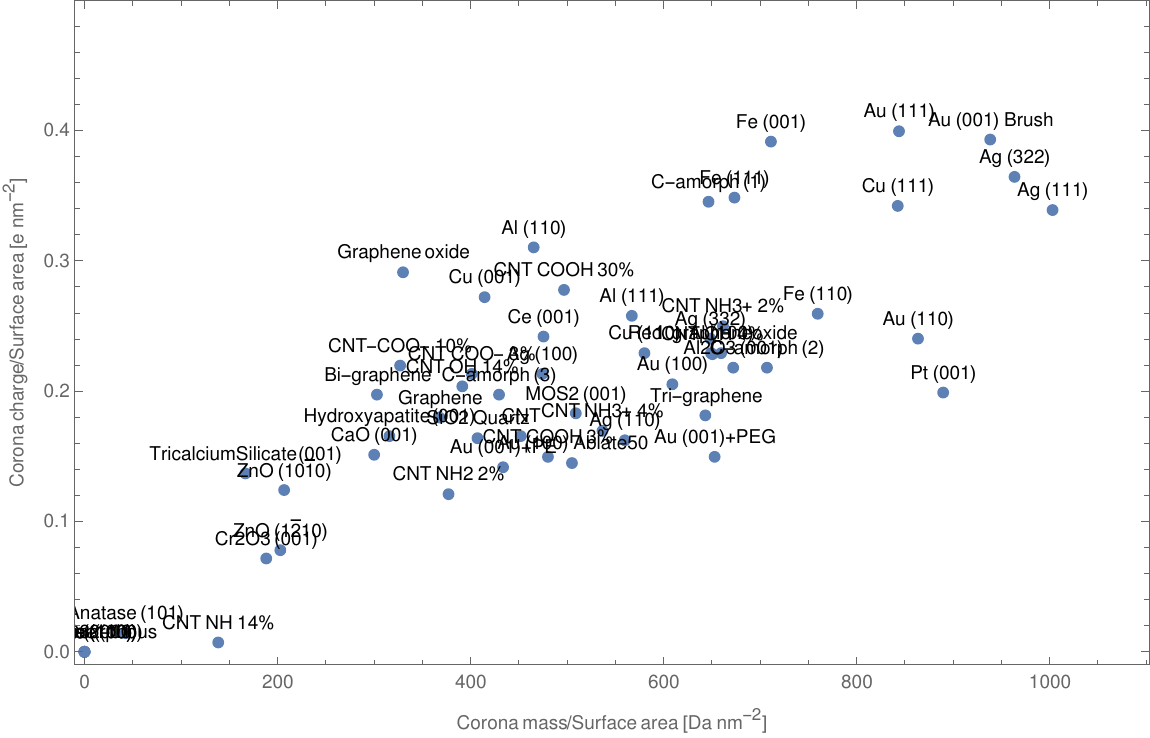}
     \caption{A plot of corona mass and charge for a variety of nanomaterials immersed in a medium of twenty proteins selected from the \emph{Daphnia magna} proteome based on a cluster analysis. }
     \label{fig:daphnia-corona}
 \end{figure}

\begin{table}[tb]
    \centering
    \begin{tabular}{cccc}
         ID&  Concentration [$\mu$M]  & Mass [kDa] & Charge [e]\\
A0A162CU06 & 3.76 & 13.31 & 4.5 \\
A0A0P5SKA7 & 1.20 & 41.56 & 12.5 \\
A0A0P5LW78 & 2.29 & 21.83 & -1.5 \\
A0A164KXJ8 & 0.48 & 103.34 & 9.5 \\
A0A0P4XG24 & 5.23 & 9.57 & 10.5 \\
A0A0N8BEG1 & 1.49 & 33.64 & -4.5 \\
A0A0P6AYL2 & 7.20 & 6.95 & 3.5 \\
A0A0P5SS69 & 2.02 & 24.80 & 3.5 \\
A0A0P5WS26 & 1.72 & 29.01 & 6.5 \\
A0A164Z8Z4 & 0.51 & 98.27 & 0.0 \\
A0A164V4J0 & 3.76 & 13.31 & 9.5 \\
A0A162NKY0 & 4.82 & 10.38 & 2.0 \\
A0A0P6BQN1 & 5.33 & 9.37 & 11.0 \\
A0A164Z4N7 & 0.79 & 63.42 & 0.0 \\
A0A164F1M8 & 1.21 & 41.37 & 8.0 \\
A0A162NFS0 & 0.87 & 57.50 & 0.0 \\
A0A164X841 & 1.37 & 36.45 & 8.0 \\
A0A164SWA3 & 8.79 & 5.69 & 4.0 \\
A0A162BQ10 & 3.55 & 14.08 & -6.0 \\
A0A164U6G0 & 1.93 & 25.90 & 5.5 \\

    \end{tabular}
    \caption{Proteins selected from the \emph{Daphnia magna} proteome based on k-means clustering of PEPSTAT and other descriptors. }
    \label{tab:daphniaProteins}
\end{table}
As a demonstration of the use of this automated scanning, we have performed corona simulations for a trial solution of twenty proteins selected from the proteome for \emph{Daphnia magna}, using the AlphaFold structures for these and selecting proteins based on clustering of their properties. These input descriptors are selected from a large set of descriptors calculated via PEPSTATS, a modification of PEPSTATS to produce properties only for surface AAs, and additional descriptors related to the structure of the protein, with the k-means algorithm used to select twenty proteins.  The resulting proteins and the concentrations assigned are given in Table \ref{tab:daphniaProteins}, where concentrations are chosen such that the mass concentration of each protein is equal. Corona predictions have been performed for a set of $70$ materials for which ML PMFs are available, which for technical reasons employed a slightly older version of the methodology but is sufficiently accurate to demonstrate the general procedure shown here. The \progname{CoronaStats.py} script included in the repository is employed to post-process these to a simple pair of descriptors per NP: the total mass and charge of adsorbates, normalised to the surface area of the NP. The resulting values are the end of the simulation (1 ms of simulated time) are plotted in Figure \ref{fig:daphnia-corona} to demonstrate the use of this pipeline in performing a rapid categorisation of nanomaterials in a given medium. It can clearly be seen that there is a clustering of metallic materials towards the right-hand side of the chart, indicating a greater total corona content while some other materials are highly charge-specific and tend to produce less abundant coronas. More generally, this technique can be employed to produce corona-total or corona-average values for a descriptor $x = \sum_i x_i N_i / \sum_i N_i$, where $x_i$ is the value of $x$ for adsorbate $i$ and $N_i$ is the number of instances of that adsorbate in the corona. This provides a convenient fingerprint of the nanomaterial with respect to a given medium for use in ML studies or categorisation of NPs. 

\section{Material Library}

PMFs and Hamaker constants have been computed and included in the repository for a wide range of materials, with a particular focus on the adsorption of AA/SCAs to these surfaces. In this section, we present an overview of the calculations used to parameterise these interactions. Calculations performed using these potentials should cite the original works. 

\subsection{PMFs}
Tabulated PMFS for sets of biomolecular fragments have been computed using metadynamics simulations for a range of materials: Au (100, 110, 111), Ag (100, 110, 111), Al, Fe, CNTs (pristine and modified with a range of functionalisations), graphene (1, 2, 3 layers), graphene oxide, reduced graphene oxide, amorphous carbon (three morphologies), titanium dioxide (two rutile, two anatase surfaces), silica (amorphous and quartz), iron oxide, cadmium selenide. Due to the differing availability of forcefields and computational methods (especially metadynamics settings, ionic strength and species, and choice of fragments) used, these PMFs differ slightly in terms of coverage of small molecules and details such as the appropriate LJ cutoff to employ. A summary of the PMFs provided in the library is presented in Table S3.  For the majority of these materials and including a range of further materials of interest, a previously developed ML approach has been employed to produce PMFs for an extended bead set \cite{rouse2023machine}. We have made some minor updates to the methodology employed, primarily a few minor adjustments to the network architecture, the inclusion of additional materials in the training set (zinc sulphide \cite{rahmani2023biomolecular}, zinc oxide \cite{saeedimasine2023ab}, FCC copper, aluminium and iron \cite{js_copper, parinazaluminium2023,parinaziron2022}), and PMFs produced for both SCAs and full AAs for rutile using an alternative forcefield \cite{brandt2015molecular,tio2pmfold}. We have also made some changes in the input potentials used, with these changes made available in the model repository \cite{Rouse_PMFPredictor-Toolkit_2024}. The most significant correction was the re-identification of a set of PMFs for the Au (100) surface as having been generated for full AAs rather than SCAs, which previously led to a decrease in the model accuracy. The final PMFs are produced ``in the style of'' the methodology used for the \ce{TiO2} and carbonaceous materials for all materials to produce a uniform standard to facilitate comparisons across materials. Likewise, we employ the same convention for the small molecule bead types to allow for comparison between nanomaterials, that is, we generate PMFs for the SCAs except for full-molecule models for proline and glycine. A naming convention for the AA beads is employed to match the standard expected by UA such that the three-letter AA codes correspond to the PMF of the SCA while the Hamaker constant is calculated for the full AA; all other beads use the same structure for both PMF and Hamaker constant. A pre-averaged HIS bead, consisting of weighted averages of the PMFs and Hamaker constants for HIE, HID and HIP, is provided for pH 7 with a range of individual charge variants for other AAs also produced. The additional materials include Ag (332) and (322), Pt (001), Ce (001), a set of weathered Au surfaces with varying percentages of surface atoms removed, Au decorated with PE and PEG brushes, hydroxyapatite (at a range of pH values and surface indices), stainless steel, tricalcium silicate, CaO, \ce{MoS}, \ce{Al2O3} and \ce{Cr2O3}, and a range of clay materials. For all of these additional surfaces, forcefields and structures were obtained using the Charmm-GUI nanomaterial modeller \cite{charmmguinanomodeller} and the InterfaceFF forcefield \cite{heinz2013thermodynamically}, with the exception for stainless steel for which the structure was obtained from Ref \cite{kanhaiya2021accurate}.

\subsection{Long-range interaction parameters}
Special attention must be paid to the computation of the parameters required to evaluate the Hamaker-like long-range potentials, these being the bead radius and Hamaker constant for each surface-bead pair. In the following sections, we discuss in more detail how these parameters have been calculated for the materials and chemicals provided in the repository. We note, however, that as with the PMFs, the user is free to supply their own calculated values. 

\subsubsection{Bead radii} \label{sec:beadradii}

When the standard bead set is used, the radius is calculated according to the methodology in \cite{alsharif2020silico}, which is calibrated to reproduce experimental protein densities from the vdW radius and coordinates of each AA. For the extended bead set which must account for arbitrary molecules, no equivalent procedure is available and we instead estimate the volume occupied by the molecule based on the LJ parameters of the atomistic representation of this molecule. We have developed two methodologies to do so. The first models each constituent atom as a sphere of radius $\sigma_i/2$ at the location given by the coordinates used in the input structure for that molecule, and generates points on the surface of each atom, with the convex-hull method used to select the points generating the outer surface of the molecule. The total volume occupied by this surface is then computed and the radii of the sphere with an equivalent volume are recorded. In the second approach, the orientation-averaged self-interaction via the LJ potential between the molecule with a copy of itself is recorded as a function of the distance between the centres of mass of the pair. We approximate that the effective radius of the bead is then one-half of the distance of the first zero-crossing at which the potential switches from repulsive to attractive, by analogy to the standard LJ potential. This latter methodology provides results which are generally consistent with the values from Ref. \cite{alsharif2020silico}, with a linear least-squares fit providing $R^2 = 0.82$ when the outlier histidine is excluded for consistency with that work. In Supplementary Table S4 we present the results for both of these methodologies compared to the values in \cite{alsharif2020silico} and those used in \cite{subbotinajpcb} and extracted from solution-phase values from Ref. \cite{counterman1999volumes}. The convex-hull radii can be seen to be an extremely good reproduction of the values found for solution phase AAs, which are likewise quite close to those found for AAs in proteins \cite{counterman1999volumes}. Thus, we recommend the use of the convex-hull radii and have supplied these in the repository. 
\subsubsection{Hamaker constants}
The Hamaker constants used as the overall energy scale for a given potential can be rigorously computed using Lifshitz theory based on optical constants for the NP and molecule \cite{israelachvili2011intermolecular}. This constant is denoted $A_{132}$, where $1$ is taken to be the AA, $2$ the NP and $3$ the medium, typically water, and is given by
the sum of a zero-frequency term $A_{132}(0)$ defined by,
\begin{equation}
    A_{132}(0) = \frac{3 \mkbt}{4} \left( \frac{ \epsilon_1 - \epsilon_3}{\epsilon_1 + \epsilon_3}\right) \left(\frac{ \epsilon_2 - \epsilon_3}{\epsilon_2 + \epsilon_3}  \right)
\end{equation}
and a high-frequency term,
\begin{equation} \label{eq:hamakerHighFreq}
    A_{132}(\nu > 0) = \frac{3 h}{4 \pi} \int_{\nu_s}^\infty \left( \frac{\epsilon_1(i \nu) - \epsilon_3(i \nu)  }{\epsilon_1(i \nu) + \epsilon_3(i \nu)} \right) \left(  \frac{\epsilon_2(i \nu) - \epsilon_3(i \nu)  }{\epsilon_2(i \nu) + \epsilon_3(i \nu)}  \right) d \nu .
\end{equation}
In the above, $\epsilon_i = \epsilon_i(0)$ is the dielectric permittivity at low frequency, while $\epsilon(i \nu)$ is the permittivity at imaginary frequencies and $\nu_s = 2 \pi \mkbt / h$. For dielectric components (solvent, non-metallic NPs, CG beads) we approximate,
\begin{equation}
    \epsilon_i(i \nu) = 1 + \frac{n_i^2 - 1}{1 + \nu^2/\nu_i^2}
\end{equation}
where $\nu_i$ is the electronic absorption frequency in the UV and $n_i$ is the refractive index at visible wavelengths. For metal  components, we instead employ,
\begin{equation}
    \epsilon_i(i \nu) = 1+ \frac{\nu_i^2}{\nu^2} ,
\end{equation}
where $\nu_i$ is the free electron gas (plasma) frequency for that metal with $\epsilon_i(0) \rightarrow \infty$ and the refractive index not defined. We assume all CG beads are dielectric and that the solvent is water, such that we only need expressions for dielectric and metallic NPs. For a dielectric NP and approximating that all three absorption frequencies are all equal to the same value denoted $\nu_e$,  $A_{132}(\nu>0)$ is approximately given by,
\begin{equation}
\label{eq:hamakerDielectric}
    A_{132} (\nu>0) \approx \frac{3h\nu_e}{8\sqrt2} 
    \frac{\left(n_1^2-n_3^2\right)\left(n_2^2-n_3^2\right)}{\sqrt{\left(n_1^2+n_3^2\right) \left(n_2^2+n_3^2\right)}\left(\sqrt{ n_1^2+n_3^2 }+\sqrt{ n_2^2+n_3^2 }\right)}.
\end{equation} 

For a metallic NP,  we numerically integrate Eq. \eqref{eq:hamakerHighFreq} using the appropriate expressions for each dielectric permittivity, allowing the absorption or plasma frequency to differ for each component.

%


These expressions are implemented in the pre-processing script \progname{CalcLifschitzHamaker.ipynb} as described earlier. Typically, the required optical constants must be found in databases of experimental results or computed from first principles, which may not be possible or be extremely time-consuming.  The estimation of Hamaker constants from first principles can be also achieved through the calculation of the LJ constant $C_6$ \cite{Hongo_2017, Takagishi_2019}. However, this method is also time-consuming and cannot be straightforwardly automated as is required here. In the case where optical data is not available, we instead approximately extract the Hamaker constant from forcefield parameters for the species in question. This methodology is implemented in the Enalos Hamaker Constant Tool (EHCT) \cite{Afantitis20206523, Varsou2020789}, which requires only empirical formulae and densities as input. An automated routine to perform the calculation is also implemented in the PMFPredictor Toolkit \cite{Rouse_PMFPredictor-Toolkit_2024}. In this method, we compute the vacuum self-interaction Hamaker constant for a given CG bead by summation over forcefield parameters similar to the method in the EHCT, 
\begin{equation}
    A = \frac{4 \pi^2 \eta^2}{V_b^2} \sum_m \sum_l \epsilon_{lm} \sigma^6_{lm}
\end{equation}
with $\epsilon_{lm}, \sigma_{lm}$ computed using combination rules, $\eta = 0.64$ is the packing density for random close-packed spheres, and $V_b$ is the approximate volume per molecule, with this volume set equal to the convex hull volume as discussed in the previous section, such that $V_b/\eta$ represents the volume occupied by the molecule in the condensed phase. This procedure typically produces values in the range $0.1 - 1.0 \times 10^{-20}$ J, in agreement with the range expected for Hamaker constants for organic molecules. We note that the computed value for water of $6.8 \times 10^{-20}$ J is slightly larger than the values presented in Ref. \cite{israelachvili2011intermolecular} of $(3.7 - 5.5)\times 10^{-20}$ J, but within an acceptable error given the generally small contribution from the Hamaker potential for beads of this size. We employ the same approach to generate Hamaker constants for each surface structure, with the exception that since the coordinates for these are assumed to already be in the solid phase we set $\eta = 1$.  In principle, combining relations can then be used to produce the Hamaker constant describing the interaction between each bead and the surface in a medium $w$,
\begin{equation}
    A_{cmw} =  \left ( \sqrt{A_{cc}} - \sqrt{A_{ww}} \right )\left ( \sqrt{A_{mm}} - \sqrt{A_{ww}} \right ),
\end{equation}
These relations, however, are known to be inaccurate when the medium is water, as is the assumed case here \cite{israelachvili2011intermolecular}. We find in particular that for metallic NPs, for which $A_{mm}$ is large, the results are highly dependent on the relative values of $A_{cc}, A_{ww}$ since this may easily produce strongly positive or negative values with slight variations in $A_{cc}$.  Thus, we instead use the Lifshitz model as described above by finding approximate values for the optical constants which are compatible with the values of $A_{cc}$ computed from forcefield parameters. We neglect the zero-frequency term since this is always less than $1 k_BT$ in magnitude. Next, we approximate that $\nu_e \approx 3\times10^{15}$ Hz for all beads, such that \eqref{eq:hamakerDielectric} can be numerically inverted to obtain $n_1 = n_2$ with $n_3 = 1$ as a function of the Hamaker constant for the chemical interacting with itself in vacuum as calculated above. We assume that $\epsilon_i$ for this bead is a nominal value of $1.3$, but note this is effectively negligible compared to that of water. Next, we identify whether the NP material should be treated as metallic or nonmetallic by analysing the fraction of its constituent atoms which are highly polarisable in terms of their forcefield parameter  $\epsilon_i > 12$  and quasi-neutral $|q| < 0.5 e$. If over half of the atoms in the structure meet this definition, the structure is taken to be metallic with a nominal plasma frequency of $5 \times 10^{15}$ Hz and $\epsilon$ set to an arbitrarily large number. Otherwise, we extract an effective refractive index through the same procedure as for CG beads and again set $\epsilon_r$ to be a nominal value of $1.3$ and assume an electronic adsorption frequency of $ 3\times10^{15}$ Hz. This produces the required set of constants for both the material and CG bead. Next, we set the medium to be water $n_3 = 1.33, \epsilon_3 = 82, \nu_w = \nu_e$, and compute Hamaker constants using the Lifshitz theory based on the extracted approximate optical constants. We stress that this is an approximate procedure and the results are generally only correct to within an order of magnitude and are dependent on the exact forcefield parameterisation used. This is particularly apparent for certain small molecules containing sp3 nitrogen in GAFF parameterisations, which has an unusually large value of $\epsilon_i$ compared to typical atoms, and in general for small molecules containing only one or two heavy atoms due to the low volumes of these beads producing an overestimated numerical density.  Given, however, the generally small contribution of the Hamaker potential to UA binding energies, this does not lead to a significant overall error, especially in the context of the limitations of the Hamaker and Lifshitz approach in general \cite{israelachvili2011intermolecular}.



\section{Methodology development and validation}
Here, we summarise the publications introducing or testing the methods described above. The CG scheme to evaluate biomolecule-NP interaction energy, the heatmaps and the ensemble average energy using the united atom one-bead-per-amino acid approach was first introduced in \cite{lopez2015coarse}. This methodology was improved to include the pre-calculated bead-NP PMFs, including the planar-to-spherical shape correction, in Ref. \cite{power2019multiscale} and implemented in the \progname{UnitedAtom} software tool in place of the EspressoMD script previously used. The method to evaluate the Hamaker constant for the interaction between NP materials and AAs using experimental AA radii and refractive indices was introduced in Ref. \cite{alsharif2020silico}. This approach was later validated with noble metal NPs. The method has shown a good correlation with experimentally measured adsorption rankings, however, the absolute binding energies were not in agreement with experimental values due to the limitations of the method (``rigid body'' model for proteins)\cite{subbotinajpcb}. To address the complex variety of available NPs (core-shell NPs, nanocomposites, layered NPs, etc.), the multicomponent ``LEGO-like'' model was introduced in Ref. \cite{subbotina2023silico} and validated by comparison to experimental results for polymer-coated NPs. The \progname{CoronaKMC} method for the prediction of protein abundances on solid surfaces was first published in Ref \cite{rouse2021hard} and the methodology for converting \progname{UnitedAtom} adsorption energies to \progname{CoronaKMC} inputs demonstrated in Ref. \cite{hasenkopf2022computational}, finding good agreement with experimental results, and further employed for modelling milk films on metallic surfaces \cite{parinaziron2022, parinazaluminium2023}. 

\section{Applications}

Our multiscale model of biomolecular corona formation on solid NPs and surfaces can be generalised to a large variety of systems. It essentially relies only on the existence of an atomistic forcefield for the target NP material which is compatible with standard forcefields for biomolecules, e.g. CHARMM or GAFF. This suggests a further integration with existing methodologies for the prediction of forcefield parameters via ML techniques is likely to be highly useful in extending the range of materials even further \cite{unke2021machine, li2017machine}.

The model can be used for the prediction of fouling of surfaces in food processing and packaging, screening materials for nanomedicine, toxicology, environmental safety, material design and medical devices. Beyond that, our model benefits from the advancements in computational tools such as PMFPredictor \cite{rouse2023machine} which makes it possible to predict potentials for arbitrary small molecules of interest such as tannic or humic acid and small metabolites for environmental safety studies or food science. Combined with the fragment-based methodology used in UA, this enables a wide range of biomolecules to be scanned across a variety of NP surfaces. 

Our current model treats biomolecules as rigid structures. Incorporating mechanisms to account for protein flexibility, such as generating an ensemble of protein structures rather than using a single structure, can improve the accuracy of the CG models and capture the interactions in diverse environments. Moreover, since one of the outcomes of the modelling is the set of preferred biomolecule orientations on the adsorbent surface, this suggests using these output orientations as starting configurations for more detailed all-atom studies of the corona or individual adsorbed proteins, which can, in turn, be used as improved structures for the inputs to NPCoronaPredict, and this procedure iterated to allow for realistic configurations with a highly optimised runtime.

The NPCoronaPredict pipeline allows for the calculation of numerical descriptors representing the properties of a range of NPs immersed in a biological fluid, but the adsorption energies of the biomolecular fragments and larger proteins are also potentially vital descriptors in categorising complex structures in a simple numerical form suitable for ML methodologies, e.g. the prediction of further interactions, functionality or safe-by-design development \cite{wyrzykowska2022representing, C8NA00142A, walkey2014protein,kamath2015,afantitis2018,xia2010index}. For these models, it has been demonstrated that simple descriptors obtained from the corona composition can correlate strongly to measures such as NP cell uptake. Here, again, we stress the importance of both the speed and flexibility of our approach to handle essentially arbitrary NP structures. This is vital to be able to provide meaningful descriptors to capture potentially subtle differences in NP structure which may yet lead to significant differences in bioactivity. The scheme we have developed can model crystalline and amorphous, organic and inorganic, modified and pristine NPs on an equal footing, avoiding the risk of requiring extensive categorical or ad-hoc descriptors to account for differences between materials. This is indispensable for fields in which only limited experimental data is available to produce these ML models to limit the potential risk of needing to discard data for materials if these do not fit into an established framework or scan the potential candidate materials even before they are produced.

\section{\label{sec:conclusions}Conclusions}
We have demonstrated an end-to-end pipeline for the prediction of the corona of adsorbates formed around an NP in a medium containing a mixture of biomolecules and other compounds. Our methodology is sufficiently flexible to allow for corona prediction for a multi-component NP immersed in media consisting of a large number of varieties of proteins and other adsorbates at a fraction of the computational time which would be required for traditional molecular dynamics simulations. All the code is available open-source for download from \cite{NPCoronaPredict-repo} together with a library of required input which covers a wide range of nanomaterials and biomolecules of interest, and further NP materials or adsorbates can be straightforwardly added by the user as required. 

\section{Software Availability}
The NPCoronaPredict package is freely available for download at https://github.com/ucd-softmatterlab. The PMFPredictor software used to produce PMFs and Hamaker constants is available from https://github.com/ijrouse/PMFPredictor-Toolkit. Both packages are provided open source. 

\section{Author contributions}
I.R,. D.P. : \progname{UnitedAtom};
I.R.: \progname{CoronaKMC}, NPCoronaPredict, NPDesigner, PMFPredictor;
I.R., J.S.: Additional pre/post-processing scripts, testing, and validation;
J.S.: parameterisation of \progname{UnitedAtom} for metallic and polymeric materials;
V.L. Funding, conceptualisation
Manuscript -- initial draft I.R. with contributions from  J.S. Revisions by all authors.
\section{Supporting Information}
Supporting Information: Additional details of UnitedAtom configuration parameters, tables of included scripts, details of included nanomaterials with MD and ML PMFs, schematics of biomolecular fragments (PDF).

\begin{acknowledgement}
We acknowledge funding from the EU Horizon2020 framework under grant agreements No. 686098 (SmartNanoTox project), No. 731032 (NanoCommons project), No. 814572 (NanoSolveIT project), and No. 101008099 (Marie Curie RISE CompSafeNano project), Horizon Europe under grant agreement No. 101092741 (nanoPASS project), and by Science Foundation Ireland through grant 16/IA/4506. We also thank all beta testers for providing useful feedback, in particular Hender Lopez.
\end{acknowledgement}

\bibliography{refs.bib}

\end{document}


\maketitle


\section{UnitedAtom configuration file}
In this section, we provide more details on the parameters required for a UnitedAtom configuration file. Table \ref{table:configOptionsSTandard} provides an overview of the parameters required for a standard run. All of these must be provided in the configuration file for the program to function including the unused options which are presently disabled but required for backwards compatibility and may be re-enabled in future versions. The exception to this rule is the ``enable-[potential]'' switches which may be removed or commented out, but in general it is not recommended or necessary to do so. In particular, disabling the surface potential is likely to lead to numerically diverging binding energies since this removes the majority of the repulsive potential at short range. The NP shape parameter is an integer defining whether the adsorption energy should be calculated for spherical (np-shape 1), cylindrical (np-shape 2, 4, 5) or cubic/planar (np-shape 3) co-ordinate systems and further specifies how input PMFs should be shape-corrected. In general, np-shapes 1,2,3 expect PMFs generated for planar surfaces while np-shapes 4,5 require PMFs generated for cylinders of diameter 1.5 nm. Note that UA does not apply error-checking to ensure input PMFs match the expected form and so caution is advised to make sure the selected shape matches the input PMF, especially for low-radius NPs for which these corrections are most significant. For planar surfaces, we recommend the use of np-shape 1 set to a larger radius, $R_{NP} > 100~$nm for reasons of numerical stability in UnitedAtom. 

\begin{table}
    \centering
    \begin{tabular}{cccc}
    Parameter & Type & Units & Notes \\
    output-directory & Path & File/Folder & \\
pdb-target  & Path & File/Folder & \\
nanoparticle-radius & Value/List & nm & \\
np-type & Integer &  & see text \\
pmf-directory  & Path &  Folder & \\
hamaker-file  & Path & File  & \\
enable-surface  & Switch &   &   \\
enable-core  & Switch &   &  \\
enable-electrostatic  & Switch &   &  \\
simulation-steps  & Value&  & unused \\ 
potential-cutoff  & Value&  & unused \\ 
potential-size & Value&  & unused \\ 
angle-delta & Value&$\circ$& unused \\ 
bjerum-length  & Value& nm & unused \\ 
debye-length  & Value& nm & \\ 
temperature  & Value & K & \\
zeta-potential =  & Value/List & V & \\
amino-acids   & List & 3-letter codes & \\
amino-acid-charges & List & Elementary charges & \\
amino-acid-radii & List & nm & \\
    \end{tabular}
        \caption{Parameters used as input in the configuration file for a standard UnitedAtom run. With the exception of the three switches, all of the above must be supplied.}
        \label{table:configOptionsSTandard}
\end{table}

In certain cases, it may be necessary to modify the standard behaviour of UnitedAtom. Table \ref{table:configOptionsExtra} contains some additional parameters which can be used to so. We recommend reading the documentation supplied with UnitedAtom for further details on these. The ``recalculate-zp'' parameter enables switching UnitedAtom to a mode in which the supplied zeta potentials are interpreted as being defined relative to a reference NP of fixed radius and shape. The electrostatic surface potentials used in the code are then calculated based on the radius and shape of the target NP to produce a constant surface charge density. The disorder-strategy parameter uses the b-factor values supplied in a PDB file to identify residues which may be disordered and degrade the performance of UA, with different disorder modes treating these residues in different ways. The ``enable-fullscan'' switch, if included in the configuration file, instructs UA to begin integration of the interaction potential such that the biomolecule's closest approach is computed to allow the NP inside hollow regions of the biomolecule. This ensures that if binding cavities are present the NP can dock into them, but may be physically unrealistic for a small NP and a hollow protein.  The ``enable-local-boltz'' switch results in the mean energies computed in a local cell of $\phi,\theta$ values to be weighted by Boltzmann factors rather than a simple average as is used by default, which typically results in a more favourable binding energy in each local cell. The option ``pdb-jitter-magnitude'', if set to a value $\sigma$ greater than zero, performs a slight perturbation of the co-ordinates of each CG bead following the generation of a random orientation, 
\begin{equation}
    x_i \rightarrow x_i + \mathcal{N}(0,\sigma)
\end{equation}
where $\mathcal{N}(\mu,\sigma)$ is a randomly generated, normally distributed variable of mean $\mu$ and standard deviation $\sigma$. Since this is applied per-axis, the total expected RMSD is equal to $\sqrt{3} \sigma$. For typical PDB resolutions of $0.2$ nm this suggests a value of $\sigma \approx 0.1$ nm. Used together with the enable-local-boltz option, this allows a very slight relaxation of the biomolecule in its local environment.

\begin{table}
    \centering
    \begin{tabular}{cccc}
    Parameter & Type & Units & Notes \\
   omega-angles & List & $\circ$ & Final rotation angles \\
   enable-fullscan & Switch & & Enables scanning to biomolecule COM  \\
   bounding-radius & Value & nm & Override bounding radius for NP \\
   overlap-penalty & Value & $k_B T$ & Extra penalty for overlap \\
   recalculate-zp & Integer & & See text \\
   calculate-mfpt & Integer & & Enable MFPT calculation \\
   disorder-strategy& Integer & & See text \\
   disorder-minbound& Float & & See text \\
   disorder-maxbound& Float & & See text \\
   enable-local-boltz & Switch &  & See text \\
   pdb-jitter-magnitude & Float & nm & Random CG bead displacement magnitude
    \end{tabular}
        \caption{Parameters used to modify UnitedAtom behaviour or enable advanced modes. All of these are optional.}
        \label{table:configOptionsExtra}
\end{table}

\section{Script library}
Here we present a brief overview of the main additional scripts supplied in the NPCoronaPredict repository:
\begin{itemize}
    \item BuildCoronaCoords.py -- Use CoronaKMC output and PDB structures to assemble the final corona for visualisation.
    \item BuildCoronaParams.py -- Convert .uam files to CoronaKMC input.
    \item ConcatPatches.py -- combines input files for CoronaKMC simulations to handle NPs with multiple crystal faces, only recommended if running separate simulations is not possible.
    \item CoronaStats.py -- Given CoronaKMC output and biomolecule descriptors, computes corona averages.
    \item ExtractBindingEnergies.py -- Scan input folders for .uam files and record average energies in a tabular format.
    \item MultiSurfaceAverage.py -- Boltzmann average adsorption energies across multiple NP surfaces.
    \item PreprocessProteins.py -- Applies PROPKA and canonical transformation to an input protein.
    \item tools/ApplyOptimumRotation.py -- Given a .uam and .pdb input, rotates the biomolecule to its favoured binding orientation for that .uam file.
    \item tools/MolToFragments.py -- Converts a SMILES code to a set of fragments and provides .pdb output for use in UA.
    \item tools/plotmap -- Generates a heatmap plot for a given .uam file.
    \item tools/CalcLifschitzHamaker -- see main text
    \item tools/ViewNP -- see main text
    item tools/VisualiseUAResults -- see main text
\end{itemize}
We note that further scripts for more specialised or testing purposes are included, but these do not form part of the core repository and may be removed at any time.
\section{Library of metadynamics PMFs}

In Table S3 we present details of the library of PMFs computed using metadynamics techniques and provide information on the sets of fragments and conditions used, with details of the suggested CG bead parameters in Table S4.

\begin{sidewaystable}
    \centering
    \begin{tabular}{cccccc}
       Material  & Force field (nano/bio) & Geometry   & CG beads  & Notes & Ref. \\
        Au (100) &  INTERFACE/GAFF &  Plane&  F & No salt & \cite{power2019multiscale} \\
        Au (100/110/111) &  INTERFACE FF/CHARMM &  Plane & S-LXG &  & \cite{subbotina2023silico}\\
        Ag (100/110/111) &  INTERFACE FF/CHARMM &  Plane & S-LXG & & \cite{subbotinajpcb}\\
        Fe (100/110/111) &  INTERFACE FF/GAFF &  Plane & SL &   & \cite{parinaziron2022}\\
        Al (100/110/111) &  INTERFACE FF/GAFF &  Plane & SL &   & \cite{parinazaluminium2023}\\
        C-amorph (3$\times$ structures) & GAFF/GAFF &  Plane & SL &   & \cite{saeedimasine2020atomistic}\\
        Graphene (1/2/3 layers) & GAFF/GAFF &  Plane & SL &  & \cite{saeedimasine2020atomistic}\\
        Graphene oxide & GAFF/GAFF &  Plane & SL&  &\cite{saeedimasine2020atomistic}\\
        Reduced graphene oxide & GAFF/GAFF &  Plane & SL&  & \cite{saeedimasine2020atomistic}\\
        CNT & GAFF/GAFF &  1.5 nm diameter cylinder & SL&  & \cite{saeedimasine2020atomistic}\\
        CNT-OH & GAFF/GAFF &  1.5 nm diameter cylinder & SL&   & \cite{saeedimasine2020atomistic}\\
        CNT-NH2 & GAFF/GAFF &  1.5 nm diameter cylinder & SL&   & \cite{saeedimasine2020atomistic}\\
        CNT-NH3+ & GAFF/GAFF &  1.5 nm diameter cylinder & SL&   & \cite{saeedimasine2020atomistic}\\
        CNT-COO- & GAFF/GAFF &  1.5 nm diameter cylinder & SL&   & \cite{saeedimasine2020atomistic}\\
        CNT-COOH & GAFF/GAFF &  1.5 nm diameter cylinder & SL &   & \cite{saeedimasine2020atomistic}\\
        Fe2O3 (001) &  A.L. /GAFF &  Plane &  SL &   & \cite{lyubartsevbionano}\\
        TiO2 anatase(100, 101) &  A. Lyubartsev /GAFF &  Plane & SL  &  & \cite{rouse2021}\\
        TiO2 rutile (100, 110) &  A. Lyubartsev/GAFF &  Plane & SL  &   & \cite{rouse2021}\\
        SiO2 (quartz, amorphous) &  INTERFACE FF/GAFF &  Plane & SL  &  & \cite{lyubartsevbionano} \\
    \end{tabular}
    \caption{ A summary of the materials included in the metadynamics PMF library as of time of publication. The CG bead section uses the following shorthand to describe available bead sets: F: full AA, S: standard SCA set, G: glycans/carboydrates, X: additional ionised variants, L: lipids and glucose. S- indicates proline is modelled as cyclopropane and glycine is absent.}
    \label{tab:materialset}
\end{sidewaystable}

\begin{sidewaystable}
    \centering
    \begin{tabular}{cccccc}
       Chemical  & r (FF-CH) [nm]  & r (LJ0) [nm]  & r (DP/SA) [nm]  & r (JS) [nm] & r(S) [nm]\\
ALA     &0.28	&0.34	&0.32 & 0.27 & 0.29 \\
ARG 	&0.37	&0.54	&0.43 & 0.35 & - \\
ASN 	&0.32	&0.35	&0.36 & 0.31 & 0.31 \\
ASP     &0.31	&0.35	&0.36 & 0.31 & 0.30 \\
CYS     &0.30	&0.32	&0.35 & 0.30 & - \\
GLN &0.34	&0.44	&0.39 & 0.32 & 0.33\\
GLU  &0.33	&0.36	&0.38 & 0.32& 0.32\\
GLY	    &0.25	&0.31	&0.29 & 0.27 & 0.26\\
HIS    & 0.34 & 	0.41 &	0.302 & 0.33 & - \\
ILE &0.34	&0.49	&0.40 & 0.31 & 0.35\\
LEU 	&0.34	&0.42	&0.40 & 0.31 & 0.35 \\
LYS 	&0.34	&0.40	&0.41 & 0.32 & 0.34 \\
MET 	&0.34	&0.41	&0.40 & 0.33 & 0.35 \\
PHE 	&0.36	&0.47	&0.42 & 0.34 & 0.36 \\
PRO	    &0.28	&0.37	&0.36 & 0.29 & 0.32\\
SER 	&0.29	&0.31	&0.33 & 0.28 & 0.29\\
THR  	&0.31	&0.36	&0.36 & 0.30 & 0.31 \\
TRP &0.38	&0.56	&0.45 & 0.37 & 0.38  \\
TYR 	&0.37	&0.49	&0.43 & 0.35 & 0.37\\
VAL &0.32	&0.40	&0.38 & 0.30 & 0.33\\
    \end{tabular}
    \caption{AA radii for Hamaker potentials computed using different methodologies: forcefield parameters and convex hull (FF-CH), zero-crossing of LJ self-interaction (LJ0), weighted average of vdW and maximum radius of gyration (DP/SA), the methodology used in \cite{subbotinajpcb} and described in \textit{CalcLifschitzHamaker.ipynb} notebook, and values extracted from solution-phase volumes (S) \cite{counterman1999volumes}.  Note that all radii here are calculated for the full molecule including the AA functional group to better reflect their behaviour in the bulk of a protein. }
    \label{tab:hamakerRadiusset}
\end{sidewaystable}

\section{Machine-learning PMFs}
\begin{figure}[tb]
     \centering
     \includegraphics[width=0.95\textwidth]{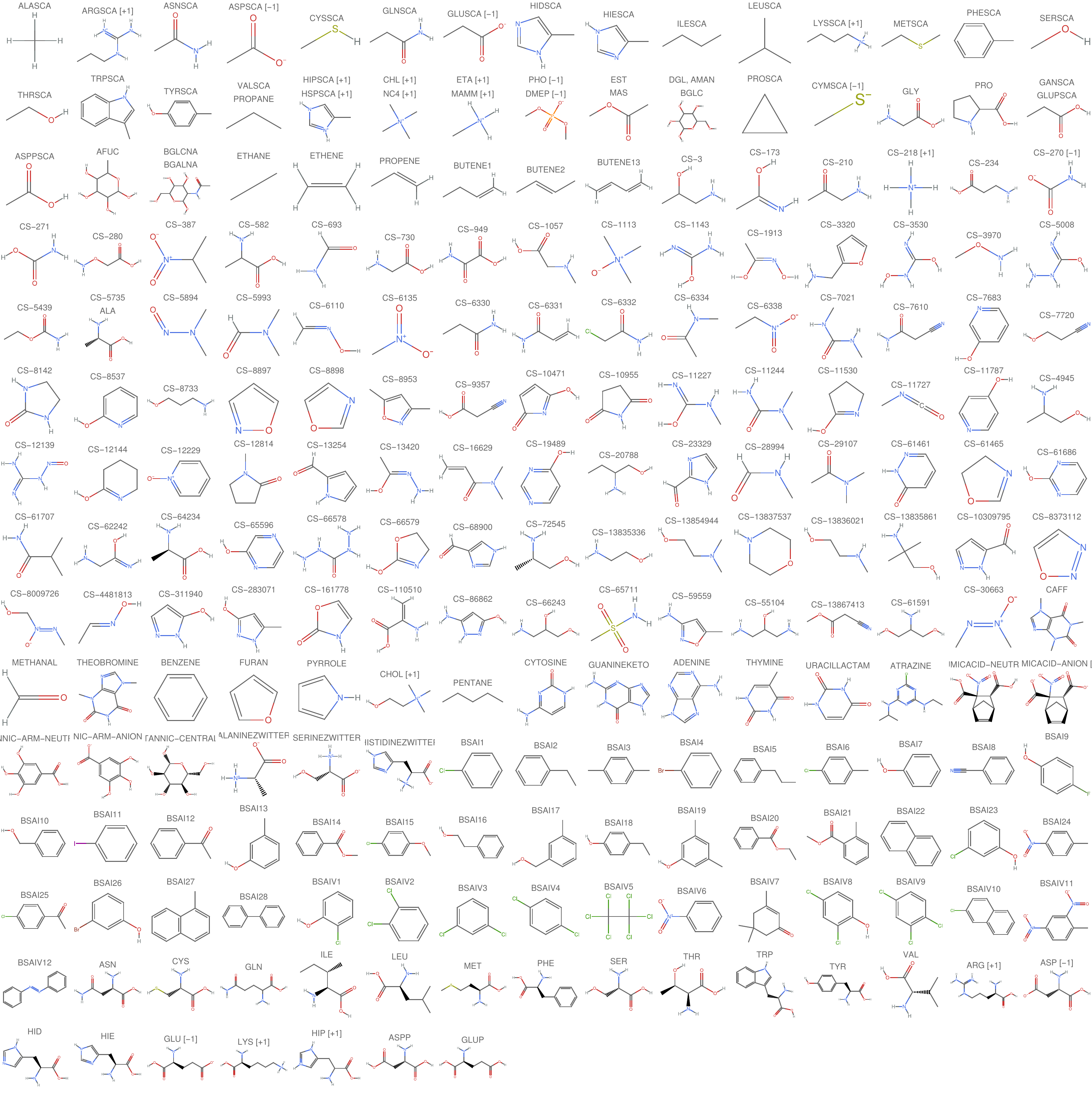}
     \caption{ Bead IDs and 2D structures for the full range of small molecules available in the extended bead set. We indicate charges where relevant after the bead ID and structures with identical SMILEs codes are merged but may have multiple bead IDs due to different naming conventions employed by different groups. }
     \label{fig:extended_beads_structures}
 \end{figure}

In this section, we provide an overview of the full set of small molecules parameterised using PMFPredictor and the associated scripts for use with UnitedAtom. Structures for these molecules are shown in Figure \ref{fig:extended_beads_structures} and detailed results (radii, Hamaker constants, SMILES codes) are provided in the NPCoronaPredict repository. Tables S7, S8 and S9 provide details of the available nanomaterials separated into metallic, carbonaceous and miscallenous groups respectively. For all of these, PMFs are generated for the full set of biomolecules and forcefields are either INTERFACEFF or the same as listed in Table S3.

\begin{table}
    \centering
    \begin{tabular}{ccccc}
   Surface ID&Shape&Type&Main elements&Notes\\
 Ag100&plane&Metal&Ag&FCC (100) \\
Ag110&plane&Metal&Ag&FCC (110) \\
Ag111&plane&Metal&Ag&FCC (111) \\
Ag322&plane&Metal&Ag&FCC (322) \\
Ag332&plane&Metal&Ag&FCC (332)\\
AlFCC100UCD&plane&Metal&Al&FCC (100) \\
AlFCC110UCD&plane&Metal&Al&FCC (110) \\
AlFCC111UCD&plane&Metal&Al&FCC (111) \\
AuFCC100&plane&Metal&Au&FCC (100) \\
AuFCC100-Ablate5&plane&Metal&Au&FCC (100) gold, 5\% ablation\\
AuFCC100-Ablate25&plane&Metal&Au&FCC (100) gold, 25\% ablation\\
AuFCC100-Ablate50&plane&Metal&Au&FCC (100) gold, 50\% ablation\\
AuFCC100-Ablate75&plane&Metal&Au&FCC (100) gold, 75\% ablation\\
AuFCC100UCD&plane&Metal&Au&FCC (100) \\
AuFCC110UCD&plane&Metal&Au&FCC (110) \\
AuFCC111UCD&plane&Metal&Au&FCC(111)\\
AuSphere1NM&sphere&Metal&Au&Spherical\\
Ce-001&plane&Metal&Ce&FCC (001) \\
Cu001&plane&Metal&Cu&FCC (001) \\
Cu110&plane&Metal&Cu&FCC (110) \\
Cu111&plane&Metal&Cu&FCC (111) \\
Fe001&plane&Metal&Fe&FCC (001) \\
Fe110&plane&Metal&Fe&FCC (110) \\
Fe111&plane&Metal&Fe&FCC (111) \\
Pt001&plane&Metal&Pt&FCC (001) \\
SS304-111&plane&Metal&Fe, Ni, Cr&Stainless steel, (111) \\
Au-001-PE&plane&Metal/organic&Au, C&PE brush\\
Au-001-PEG&plane&Metal/organic&Au, C&PEG brush\\
GoldBrush&plane&Metal/organic&Au, C&Gold with partial polymer brush\\

    \end{tabular}
        \caption{Table of metallic nanomaterials with PMFs generated via the PMFPredictor method.}
        \label{table:PMFPMetals}
\end{table}

\begin{table}
    \centering
    \begin{tabular}{ccccc}
   Surface ID&Shape&Type&Main elements&Notes\\
 Al2O3-001&plane&Metal oxide&Al, O&(001) \\
CaO001&plane&Metal oxide&Ca, O&(001) \\
Cr2O3-001&plane&Metal oxide&Cr, O&(001) s\\
Fe2O3-001O&plane&Metal oxide&Fe, O&(001)  hydroxylated\\
Rutile-100-old&plane&Metal oxide&Ti, O&(100) \\
SiO2-Amorphous&plane&Metal oxide&Si, O&Amorphous silica\\
SiO2-Quartz&plane&Metal oxide&Si, O&Quartz\\
TiO2-ana-100&plane&Metal oxide&Ti, O&Anatase, (100)\\
TiO2-ana-101&plane&Metal oxide&Ti, O&Anatase, (101)\\
TiO2-rut-100&plane&Metal oxide&Ti, O&Rutile, (100)\\
TiO2-rut-110&plane&Metal oxide&Ti, O&Rutile, (110)\\
ZnO10m10BOND&plane&Metal oxide&Zn, O&Bonded water model, (10-10) \\
ZnO10m10NB&plane&Metal oxide&Zn, O&Nonbonded water model, (10-10) \\
ZnO10m10NBDry&plane&Metal oxide&Zn, O&Nonbonded, dehydrated, (10-10) \\
ZnO1m210BOND&plane&Metal oxide&Zn, O&Bonded water model, (1-210) \\
ZnO1m210NB&plane&Metal oxide&Zn, O&Nonbonded water model, (1-210) \\
ZnO1m210NBDry&plane&Metal oxide&Zn, O&Nonbonded, dehydrated, (1-210) \\
mos2-001&plane&Metal sulphide&Mo, S&(001) \\
ZnS-NP&sphere&Metal sulphide&Zn, S&Spherical\\
ZnS-sphalerite-110&plane&Metal sulphide&Zn, S&(110)\\
ZnS-sphalerite-110-coated&plane&Metal sulphide&Zn, S, C&(110), dense polymer brush\\

    \end{tabular}
        \caption{Table of metal oxides and sulphides with PMFs generated via the PMFPredictor method.}
        \label{table:PMFPMetals}
\end{table}

\begin{table}
    \centering
    \begin{tabular}{ccccc}
   Surface ID&Shape&Type&Main elements&Notes\\
graphene&plane&Carbonaceous&C&Graphene (one layer)\\
bi-graphene&plane&Carbonaceous&C&Graphene (two layers)\\
tri-graphene&plane&Carbonaceous&C&Graphene (three layers)\\
grapheneoxide&plane&Carbonaceous&C, O&Graphene oxide\\
redgrapheneoxide&plane&Carbonaceous&C, O&Reduced graphene oxide\\
C-amorph-1&plane&Carbonaceous&C&Amorphous carbon\\
C-amorph-2&plane&Carbonaceous&C&Amorphous carbon\\
C-amorph-3&plane&Carbonaceous&C&Amorphous carbon\\
CNT15&cylinder&Carbonaceous&C&CNT\\
CNT15-COO--10&cylinder&Carbonaceous&C, O&CNT, COO- 10\%\\
CNT15-COO--3&cylinder&Carbonaceous&C, O&CNT, COO- 3\%\\
CNT15-COOH-3&cylinder&Carbonaceous&C, O&CNT, COOH 3\%\\
CNT15-COOH-30&cylinder&Carbonaceous&C, O&CNT, COOH 30\%\\
CNT15-NH2-14&cylinder&Carbonaceous&C, N&CNT, NH2 14\%\\
CNT15-NH2-2&cylinder&Carbonaceous&C, N&CNT, NH2 2\%\\
CNT15-NH3+-2&cylinder&Carbonaceous&C, N&CNT, NH3+ 2\%\\
CNT15-NH3+-4&cylinder&Carbonaceous&C, N&CNT, NH3+ 4\%\\
CNT15-OH-14&cylinder&Carbonaceous&C, O&CNT, OH 14\%\\
CNT15-OH-4&cylinder&Carbonaceous&C, O&CNT, OH 4\%\\

    \end{tabular}
        \caption{Table of carbonaceous nanomaterials with PMFs generated via the PMFPredictor method.}
        \label{table:PMFPMetals}
\end{table}

\begin{table}
    \centering
    \begin{tabular}{ccccc}
   Surface ID&Shape&Type&Main elements&Notes\\
 Gypsum-001&plane&Mineral&Ca, S, O&(001)\\
Gypsum-010&plane&Mineral&Ca, S, O&(000) \\
Gypsum-011&plane&Mineral&Ca, S, O&(011) \\
hydroxyapatite-001-ph5&plane&Mineral&Ca, P, O&(001), pH5\\
hydroxyapatite-001-ph10&plane&Mineral&Ca, P, O&(001), pH10\\
hydroxyapatite-001-ph14&plane&Mineral&Ca, P, O&(001), pH14\\
hydroxyapatite-010-ph5&plane&Mineral&Ca, P, O&(010), pH5\\
hydroxyapatite-010-ph10&plane&Mineral&Ca, P, O&(010), pH10\\
hydroxyapatite-010-ph14&plane&Mineral&Ca, P, O&(010), pH14\\
hydroxyapatite-101-ph5&plane&Mineral&Ca, P, O&(101), pH5\\
hydroxyapatite-101-ph10&plane&Mineral&Ca, P, O&(101), pH10\\
hydroxyapatite-101-ph14&plane&Mineral&Ca, P, O&(101), pH14\\
Kaolinite-001&plane&Mineral&Al, S, O&(001) \\
LithiumCobaltOxide-001&plane&Mineral&Li, Co, O&(001)\\
Montmorillonite-001-CEC0&plane&Mineral&Si, Al, Mg, O&(001)\\
Montmorillonite-001-NaCEC87&plane&Mineral&Si, Al, Mg, O, Na&87mmol/100g Na\\
Montmorillonite-001-NaCEC90&plane&Mineral&Si, Al, Mg, O, Na&90mmol/100g Na\\
Montmorillonite-001-NaCEC108&plane&Mineral&Si, Al, Mg, O, Na&108mmol/100g Na\\
Montmorillonite-001-NaCEC143&plane&Mineral&Si, Al, Mg, O, Na&143mmol/100g Na\\
Muscovite-001&plane&Mineral&K, Al, Si&(001) \\
Pyrophyllite-001&plane&Mineral&Al, Si, O&(001) \\
Tobermorite-004&plane&Mineral&Ca, Si, O&(001) \\
TricalciumAluminate-Wet-010&plane&Mineral&Ca, Al, O&(004)\\
TricalciumSilicate-Wet-001&plane&Mineral&Ca, Si, O&(010), hydated\\
TricalciumSilicate-Wet-010&plane&Mineral&Ca, Si, O&(010), hydated\\
TricalciumSilicate-Wet-100&plane&Mineral&Ca, Si, O&(100), hydrated\\
TricalciumSilicate001&plane&Mineral&Ca, Si, O&(001)\\
TungstenDisulfide-001&plane&Mineral&W, S&(001) \\
CdSeWurtzite2-10&plane&Semiconductor&Cd, Se&(2-10) \\

    \end{tabular}
        \caption{Table of other nanomaterials with PMFs generated via the PMFPredictor method. All Montmorillionite PMFs correspond to the (001) surface, with this label ommitted for space. }
        \label{table:PMFPMetals}
\end{table}

\bibliography{refs.bib}